\documentclass[12pt]{article} 
\usepackage{latexsym,amsmath,amsfonts,amsthm,amssymb}

\usepackage{graphics}
\usepackage[usenames]{color}

\newcommand{\cC}{\mathcal{C}}
\newcommand{\cD}{\mathcal{D}}

\newcommand{\cG}{\mathcal{G}}
\newcommand{\cH}{\mathcal{H}}

\newcommand{\cL}{\mathcal{L}}

\newcommand{\cS}{\mathcal{S}}

\newcommand{\CC}{\mathbb{C}}
\newcommand{\RR}{\mathbb{R}}

\newcommand{\one}{1\!\!\!1}

\newcommand{\EE}[1]{\mathbb{E}_{0}\left(#1\right)}
\newcommand{\EEa}[1]{\mathbb{E}_{ a}\left(#1\right)}
\newcommand{\pair}[1]{\langle#1\rangle}

\newcommand{\psibar}{\bar{\psi}}

\newtheorem{theorem}{Theorem}

\newtheorem{proposition}[theorem]{Proposition}
\newtheorem{corollary}[theorem]{Corollary}
\newtheorem{definition}[theorem]{Definition}
\newtheorem{lemma}[theorem]{Lemma}

\newcommand{\remark}{\medskip \noindent \emph{Remark}\ \ }

\newcommand{\join}{\stackrel{\leftrightarrow}{\Delta}}

\renewcommand{\Re}{\mathrm{Re}\,}

\newcommand{\bhat}{\hat{\beta }} 
\newcommand{\effbeta}[1]{\beta _{\mathrm{eff}, #1}}
\newcommand{\phibar}{\bar{\phi }}
\newcommand{\cB}{\mathcal{B}}
\newcommand{\Db}{\cD _{\beta}}
\newcommand{\Dbarb}{\overline{\cD} _{\beta }}
\newcommand{\Dlambda}{\cD _{\lambda}}
\newcommand{\Dbarlambda}{\overline{\cD} _{\lambda}}

\newcommand{\ie}{\emph{i.e., }}
\newcommand{\eg}{\emph{e.g., }}

\newcommand{\Proof}[1]{\emph{{ Proof}#1.}}

\begin{document}

\title{Green's Function for a Hierarchical Self-Avoiding Walk in
Four Dimensions}
\author{
David C. Brydges\thanks{Research supported by NSF Grant DMS-9706166}\\
  	University of British Columbia\\
	Mathematics Department\\
	\#121-1984 Mathematics Road \\ 
	Vancouver, B.C. V6T 1Z2\\  
	Canada\\
{\tt db5d@math.ubc.ca}\\[1mm] 
and\\[1mm]
Department of Mathematics \\ 
Kerchof Hall \\ 
P. O. Box 400137 \\ 
University of Virginia\\[3mm] 
\and
John Z. Imbrie \\
Department of Mathematics \\ 
Kerchof Hall \\ 
P. O. Box 400137 \\ 
University of Virginia\\ 
Charlottesville, VA 22904-4137\\
{\tt ji2k@virginia.edu}
}

\date{}
\maketitle
\thispagestyle{empty}

\begin{abstract} This is the second of two papers on the end-to-end
distance of a weakly self-repelling walk on a four dimensional
hierarchical lattice. It completes the proof that the expected
value grows as a constant times $\sqrt{T}\log^{\frac{1}{8}}T
\left(1+O\left(\frac{\log\log T}{\log T}\right) \right)$, which is the
same law as has been conjectured for self-avoiding walks on the simple
cubic lattice $\mathbb{Z}^4$.

Apart from completing the program in the first paper, the main result
is that the Green's function is almost equal to the Green's function
for the Markov process with no self-repulsion, but at a different
value of the killing rate $\beta $ which can be accurately calculated
when the interaction is small. Furthermore, the Green's function is
analytic in $\beta $ in a sector in the complex plane with opening
angle greater than $\pi $.
\end{abstract}

\newpage
\pagenumbering{roman}

\tableofcontents

\newpage

\pagenumbering{arabic}
\setcounter{page}{1}

\section{Introduction}\label{section:introduction}
\setcounter{theorem}{0}
\setcounter{equation}{0}

This paper is the second in a series of two papers in which we study
the asymptotic end-to-end distance of weakly self-avoiding walk on the
\emph{four dimensional Hierarchical} lattice $\cG $. The reader is
referred to \cite{BEI92} or the first paper \cite{BrIm01a}, henceforth
referred to as paper I, for definitions of these terms.  Results from
the first paper have the prefix ``I''.  

In paper I we proved that the self-avoidance causes a $\log ^{1/8}T$
correction in the expected end-to-end distance after time $T$ relative
to the $\sqrt{T}$ law of a simple random walk.  Paper I was devoted to
the problem of how to recover the end-to-end distribution by taking
the inverse Laplace transform of the Green's function, assuming that
the Green's function has certain properties,  which are proved in
this paper in Theorem~\ref{theorem:main} and
Proposition~\ref{prop:beta-lambda-recursion}. These properties are
of independent interest and, with minor changes, should also hold for
the Green's function for the simple cubic four dimensional lattice. We
prove they hold for the hierarchical problem in this paper.

The interacting \emph{Green's function} is defined by the Laplace
transform
\begin{equation}\label{eq:intro-1}
G_{\lambda } (\beta ,x) = 
\int _{0}^{\infty } e^{-\beta T}
\EE{\one _{{\omega(T)}=x} e^{-\lambda \int _{\cG }\tau ^{2}_{x}\,dx}} \, dT ,
\end{equation}
where $\tau_{x} = \tau^{(T)}_{x}$ is the time up to $T$ that ${\omega(t)}$
is at site $x$. Our main result is Theorem~\ref{theorem:main}.  It
says that the interacting Green's function is almost equal to the
Green's function for the Markov process, $\lambda =0$, but at a
different value of the $\beta $ parameter which depends on $x$ and
$(\beta , \lambda )$ and which can be accurately calculated when
$\lambda $ is small.  The error in the approximation decays more
rapidly than the Green's function because it contains, as a prefactor,
the ``running coupling constant'' $\lambda _{N (x)}$. Furthermore the
Green's function is analytic in $\beta $ in a sector in the complex
plane with opening angle greater than $\pi $.  This very large domain
of analyticity seems to be needed for accurately inverting the Laplace
transform to calculate the end-to-end distance.

The main theorem refers to domains
\begin{eqnarray}\label{eq:intro-2}
\Db & = & \{\beta \neq 0: \ |\arg \beta| < b_\beta\};
\nonumber \\[4mm] 
\Dlambda & = & \{\lambda: \ 0 < |\lambda| <
\delta \mbox{ and } |\arg \lambda| < b_\lambda\}.
\end{eqnarray}
 For details see paper I, but, for example, we can choose
$(b_\beta,b_\lambda) =
\left(\frac{5\pi}{8},\frac{\pi}{8}\right)$.  The main theorem also
refers to a recursion: given $(\beta _{0},\lambda _{0})$ we define the
Renormalization Group (RG) recursion $(\beta _{j},\lambda _{j})$ in
Sections~\ref{section:RenormalizationTransformations} and
\ref{section:coordinates}  and establish the recursive properties
in Proposition~\ref{prop:beta-lambda-recursion}.   Having
established these estimates, we know from paper I that the recursion
has various properties.  In particular, from Proposition I.1.3, for
each $\lambda _{0}$ in the domain there exists a special choice $\beta
_{0} = \beta ^{c} (\lambda _{0})$ for the initial $\beta $ such that
the RG recursion $(\beta ^{c}_{n},\lambda _{n})$ is defined for all
$n$ and $\beta ^{c}_{n} \rightarrow 0$.  This should be viewed as a
partial description of a stable manifold for the fixed point $(0,0)$,
but note that our RG recursion is not autonomous because there are
other degrees of freedom, for example, the $r$ in
Section~\ref{section:coordinates}, which have been  projected 
out in this simplified description.  We called $(\beta
^{c}_{n},\lambda _{n})$ the \emph{critical trajectory}.  For $\beta
_{0}$ some other choice of initial data, we defined the deviation
$\bhat _{n} : = \beta _{n} - \beta ^{c}_{n}$ of its trajectory from
the critical trajectory. The main result is

\begin{theorem}\label{theorem:main}
{ Let $\lambda_0 \in \Dlambda$ with $\delta$ sufficiently small. Then }
$G_{\lambda _{0}}(\beta
_{0},x)$ is analytic in $\beta _{0}$ in the domain $\cD_\beta
+\beta^c(\lambda _{0})$ and
\begin{equation}\label{eq:1.18}
|G_{\lambda _{0}}(\beta _{0},x)-G_0(\beta_{{\rm eff},N(x)},x)| 
\leq O(\lambda_{N(x)})|G_0(\beta_{{\rm eff},N(x)},x)|.
\end{equation}
Here $N(x) = \log |x|$ for $ x \neq 0$, $N(0) = 0$, and $\beta_{{\rm
eff},j} = L^{-2j} \hat{\beta}_j.$
\end{theorem}
This theorem and Proposition~\ref{prop:beta-lambda-recursion} are the
two results needed to complete the results in paper I.

The paper begins in Section~\ref{section:tau-isomorphism} with a
review of an isomorphism that recasts the Green's function as an
almost Gaussian integral with very special properties (supersymmetry).
The virtue of this representation is that it leads to a precise
definition of a Renormalization Group (RG) transformation which
relates the Green's function with given interaction to a rescaled
Green's function with smaller interaction.  The RG transformation is
defined in Section~\ref{section:RenormalizationTransformations} and
its effect on the interaction is further described in
Section~\ref{section:coordinates}. To use this RG transformation we
need an approximate calculation of its effect on the interaction.
This is carried out in
Section~\ref{section:second-order-perturbation-theoryI}.   The
proof of the main Theorem~\ref{theorem:main} is in Section 6.  All the
analysis in this paper is in Section~\ref{section:norms}.   The
methods introduced there have several noteworthy features: (i) the
demonstration that analysis with supersymmetric integrals containing
differential forms $\equiv$ Fermions is { possible }; (ii) the use of
rotation of contours of integration in the supersymmetric integral to
obtain the large domain of analyticity needed for inverting the
Laplace transform; and (iii) simultaneous control over behavior in $x$
and behavior in $\beta $ both large and small.

With Steven Evans we wrote an earlier paper \cite{BEI92} on this same
model which studied the Green's function but only at the critical
value of the killing rate.  Here we have opted for some repetition of
ideas in that paper because we have since learned that the Grassmann
algebras used in \cite{BEI92} are natural differential forms and we
wanted to incorporate this insight systematically.

\section{$\tau $ Isomorphism} \label{section:tau-isomorphism}
\setcounter{theorem}{0}
\setcounter{equation}{0}

In a precise sense that we will now review, the differential form
\[
\phi_{x}\phibar _{x} + \frac{1}{2\pi i}d\phi_{x} \, d\phibar _{x} 
\]
represents the time $\tau _{x}$ a finite state Markov process occupies
state $x$. The following is a distillation of ideas in papers
\cite{McK80,PaSo80,Lut83,LeJ87}. 

The forms $d\phi_{x}, d\phibar _{x}$ are multiplied by the
wedge product.  To connect with notation in \cite{BEI92,BrMu91}, we set
\[
	\psi_{x} = (2\pi i)^{-1/2} d\phi _{x}, \ 
	\psibar_{x} = (2\pi i)^{-1/2} d\phibar _{x} ,  
\]
 where $(2\pi i)^{-\frac{1}{2}}$ is a fixed choice of square
root. Let $\Lambda $ be a finite set.  Given any matrix $A_{xy}$
with indices $x,y \in \Lambda $ we define the even differential form
\begin{equation}\label{eq:S-sub-A}
	S_{A} 
=
	\sum \phi_{x}A_{xy}\phibar _{y} + 
	\sum  \ \psi_{x} \, A_{xy} \, \psibar_{y},
\end{equation}
and then the exponential of this form is defined by the Taylor series
\begin{align*}
	e^{-S_{A}}
&= 
	e^{-\sum \phi_{x}A_{xy}\phibar _{y} }
	\sum \frac{1}{n!} (-\sum\psi_{x}\, A_{xy} \, \psibar_{y})^{n}.
\end{align*}
The series terminates after finitely many terms because the
anticommutative wedge product vanishes if the degree of the form
exceeds the real dimension $2|\Lambda |$ of $\CC ^{\Lambda}$.

By \emph{definition} $\int_{\CC ^{\Lambda}} $ vanishes
on forms that are of degree less than the real dimension of $\CC
^{\Lambda }$. For example, taking $|\Lambda |=1$ and $A_{xy} = A>0$,
we find that $\int_{\CC} \ e^{-S_{A}}=1$ because only the $n=1$ term
in the expansion of the exponential contributes and this term is
\begin{align*}
-(2\pi i)^{-1} A \int_{\CC} \ e^{-\phi A \phibar }
\,d\phi \,d\phibar 
= 
\frac{A}{\pi } \iint \ e^{- A (u^{2}+v^{2})} \,du\,dv = 
1,
\end{align*}
using $\phi = u+iv$ and $d\phi\, d\phibar = -2i\,du dv$. Thus
these integrals are self-normalizing.  This feature generalizes: 

\begin{lemma} \label{lem:gaussian1} Suppose that $A$ has positive real
part, meaning $\Re \sum \phi_{x}A_{xy}\phibar _{y} > 0$ for $\phi \not
=0$. Then
\begin{align*}
&
\int_{\CC ^{\Lambda}} \ e^{-S_{A}} = 1,\\
&
\int_{\CC ^{\Lambda}} \ e^{-S_{A}}\ \phi_{a}\phibar _{b} =
C_{ab} ,
\end{align*}
where $C = A^{-1}$.
\end{lemma}

The second part of the lemma follows from the first part together with
the standard fact that the covariance of a normalized Gaussian measure
is the inverse of the matrix in the exponent.  The first part is a
corollary of Lemma~\ref{lem:gaussian2} given below.

Let
\[
\tau_{x} = \phi_{x} \phibar _{x} + \psi_{x} \, \psibar_{x} ,
\]
and let $\tau $ be the collection $(\tau_{x})_{x\in \Lambda }$.  Given
any smooth function $F (t)$ defined on $\RR ^{\Lambda}$, we use the
terminating Taylor series
\[
	F (\tau )
=
	\sum _{\alpha }\frac{1}{\alpha !}F^{(\alpha )}
	(\phi\phibar )(\psi \psibar)^{\alpha}
\]
to define the form $F (\tau )$, where $\phi\phibar = (\phi_{x}\phibar
_{x})_{x\in \Lambda }$, $(\psi \psibar)^{\alpha } = \prod (\psi_{x}
\psibar_{x})^{\alpha _{x}}$. The even degree of $\tau _{x}$ relieves
us of any necessity to specify an order for the product over forms.

\emph{Supersymmetry:} There is a flow on $\CC ^{\Lambda}$ given by $\phi_{x}
\longmapsto \exp (-2\pi it )\phi_{x} $. This flow is generated by the
vector field $X$ such that $X (\phi _{x}) = -2\pi i \phi _{x}$ and $X
(\phibar _{x}) = 2\pi i \phibar _{x}$. A form $\omega$ is
\emph{invariant} (under this flow) if it is unchanged by the
substitution $\phi_{x} \longmapsto \exp (-2\pi it )\phi_{x} $. The Lie
derivative $\cL _{X}\omega$ of a form $\omega $ is obtained by
differentiating with respect to the flow at $t=0$ so invariance is
equivalent to $\cL_X\omega =0$.

Let $i_{X}$ be the interior product with the vector field $X$. The
\emph{supersymmetry} operator \cite{Wit92,AtBo84}
\begin{equation}\label{eq:Q}
Q = d + i_{X}
\end{equation}
is an anti-derivation on forms with the property that $Q^{2} = d i_X +
i_{X}d = \cL _{X}$ is the Lie derivative.  Therefore $Q^{2} =0$ on
forms $\omega $ which are invariant.  A form $\omega$ that satisfies
the stronger property $Q\omega=0$ is said to be
\emph{supersymmetric}. For example $S_{A}$ is supersymmetric and the
derivation property implies that $\exp (-S_{A} )$ is also
supersymmetric.  Since $Q\phi _{x}d\phibar _{x} = d\phi _{x}d\phibar
_{x}+ (2\pi i)\phi _{x}\phibar _{x}$, there is a form $u_{x}$ such
that $\tau _{x}=Qu_{x}$.  $S_{A}$ is also in the image of $Q$.

For any form $u$ whose coefficients decay sufficiently rapidly at
infinity,
\[
\int_{\CC ^{\Lambda}} Qu =0 ,
\]
because the integral of $du$ is zero by Stokes theorem, while the
integral of $i_{X}u$ is zero because it cannot contain a form of
degree $2|\Lambda |$.  

Let $F \in \cC _{0}^{\infty } (\RR ^{\Lambda })$.  Then $\int e^{-S_{A}} F
(\lambda \tau ) $ is independent of $\lambda $, because the $\lambda $
derivative has the form
\[
\int_{\CC ^{\Lambda}} e^{-S_{A}} \sum _{x} F_{t_{x}} (\lambda \tau )
Qu_{x} = \int_{\CC ^{\Lambda}} Q \bigg(e^{-S_{A}} \sum _{x} F_{t_{x}}
(\lambda \tau )u_{x}\bigg) = 0.
\]
The compact support condition is a simple way to be sure that there
are no boundary terms at infinity. Adequate decay of the integrand
and its partial derivatives is all that is needed. If the exponential
has better decay then there is no need for such a strong condition on
$F$.  Thus

\begin{lemma} \label{lem:gaussian2} If $A$ has positive real part
and $F$ is smooth on $\RR ^{\Lambda}$ with bounded
derivatives, then
\[
\int_{\CC ^{\Lambda}} \ e^{-S_{A}} \ F (\tau ) = F (0).
\]
\end{lemma}
Part (1) of Lemma~\ref{lem:gaussian1} is obtained when $F=1$. 

The following Proposition will be called the \emph{$\tau $
isomorphism}.  It is the main result of this section.

\begin{proposition} \label{prop:parisi-sourlas}\cite{PaSo79,McK80}
Suppose that $A$ generates a Markov process with killing on first exit
from $\Lambda $.  { Let $\EEa{\cdot}$ denote the associated expectation over paths $\omega(t)$ such that $\omega(0)=a$.} Let $F \in \cC _{0}^{\infty } (\RR ^{\Lambda
})$, then
\begin{align*}
& \int_{\CC ^{\Lambda}} \, e^{-S_{A}} \, F (\tau ) \,
\phi_{a}\phibar _{b} 
= \int _{0}^{\infty } \, dT \,  \EEa{F (\tau) \,
\one _{ \omega(T) =b}} .
\end{align*}
On the right-hand side $\tau_{x} = \tau_{x}^{T}$ is the time up to $T$
that the stochastic process {$\omega(t)$} is at site $x$,
\[
\tau _{x}^{T} = \int_{0}^{T}\one _{{\omega(t)}=x}\,dt.
\]
On the left-hand side it is the form $\phi _{x} \phibar _{x}+ \psi
_{x} \psibar _{x}$.
\end{proposition}

As noted above, the compact support condition on $F$ is stronger than
necessary.  The left-hand side is a linear combination of integrals
that involve finitely many derivatives of $F$. It is still valid if
these derivatives have adequate decay at infinity, as will be the case
for functions used in this paper.

\proof{} We can assume, with no loss of generality, that $\Re \sum
\phi_{x}A_{xy}\phibar _{y} > 0$ when $\phi \not =0$, because both
sides are unchanged by
\begin{equation}\label{eq:shift-by-kappa}
F (t) \mapsto F (t) e^{\kappa \sum t_{x} }, \hspace{2mm} A \mapsto
A+\kappa I.
\end{equation}
Consider the special case where $F (\tau ) = \exp (i\sum
k_{x}\tau_{x})$.  Then
\begin{equation}\label{eq:proof-tau-isomorphism}
\int_{\CC ^{\Lambda}} \ e^{-S_{A}} \ F (\tau ) \
\phi_{a}\phibar _{b} 
= 
\int_{\CC ^{\Lambda}} \ e^{-S_{A-iK}} \phi_{a}\phibar _{b} ,
\end{equation}
where $K$ is the diagonal matrix $k_{x}\delta _{xy}$. By
Lemma~\ref{lem:gaussian1} the right-hand side equals
$(A-iK)^{-1}_{ab}$ which is 
\[
\int _{0}^{\infty } (e^{-T[A-iK]})_{ab}=
\int _{0}^{\infty } \ dT \  \EEa{F (\tau) \
\one _{\omega(T)=b}} .
\]
by the Feynman-Kac formula\footnote{ The Feynman-Kac formula is given
in \cite{Sim79} for Brownian motion. The proof that uses the Trotter
product formula is valid for finite state Markov processes.} and $F
(\tau ) = \exp (i\sum k_{x}\tau_{x}) = \exp (i\int _{0}^{T}
k_{ \omega(s) }\, ds)$.

By (\ref{eq:proof-tau-isomorphism}) the proposition is proven for $F
(\tau ) =\exp (i\sum k_{x}\tau_{x})$.  Both sides of the proposition
are linear in $F$ so we can generalize to $F \in \cC _{0}^{\infty }$
by substituting the Fourier inversion formula
\[
F (\tau ) = (2\pi )^{-n} \int \hat{F} (k)e^{i\sum k_{x}\tau_{x}}
d^{|\Lambda| }k 
\]
into $\int \exp (-S_A)F (\tau )\phi _{a}\phibar _{b}$. Since the $k$ and
$\phi $ integrals, by (\ref{eq:shift-by-kappa}), are absolutely
convergent, the integral over $k$ may be interchanged with the $\phi $
integrals. \qed


\section{Renormalization Transformations}
\label{section:RenormalizationTransformations}

\setcounter{theorem}{0}
\setcounter{equation}{0} 

\newcommand{\T}{\mathbb{T}_{\beta }}
\renewcommand{\EE}[1]{\mathbb{E}^{\Lambda }\left(#1\right)}
\newcommand{\EEo}[1]{\mathbb{E}_0^{\Lambda }\left(#1\right)}

Since the $\tau $ isomorphism is only applicable when the state space
is finite we have to study the Green's function as a limit of processes
with finite state spaces.

Let $N$ be a positive integer and let $\EEo{\cdot}$ be the expectation
for the hierarchical Levy process ${\omega(t)}$ killed on first exit from
$\Lambda =\cG _{N}$. 
Define the \emph{finite volume} interacting Green's function
\begin{equation}\label{eq:dirichlet-1}
G_{\lambda }^{\Lambda} (\beta ,x) = 
\int _{0}^{\infty } e^{-\beta T}
\EEo{\one _{{\omega(T)}=x} e^{-\lambda \int _{\cG}\tau ^{2}_{x}\,dx}} \,dT.
\end{equation}

When $\lambda =0$, $G_{\lambda =0}^{\Lambda } (\beta ,x)$ is the
$\beta $ potential for the hierarchical Levy process ${\omega(t)}$ killed on
first exit from $\Lambda =\cG _{N}$.  In this section we single out
this important object with the notation $U^{\Lambda } (\beta
,x)=G_{\lambda =0}^{\Lambda } (\beta ,x)$.  

Given a bounded smooth function $g (t)$ we define a generalization
of the Green's function
\begin{equation}\label{eq:rg-trans1}
G_{g}^{\Lambda } (\beta ,x) =
\int _{0}^{\infty } e^{-\beta T}
\EEo{ g^{\Lambda }\one _{{\omega(T)}=x}} \,dT ,
\end{equation}
where $g^{\Lambda } = \prod _{x \in \Lambda }g (\tau _{x})$.  When $g$
is the function $g (t) = \exp (-\lambda t ^{2} )$ this is the Green's
function (\ref{eq:dirichlet-1}). 
By the $\tau $ isomorphism,
\begin{equation}\label{eq:tau-with-g}
G_{g}^{\Lambda } (\beta ,x) = \int _{\CC ^{\Lambda }} \mu _{\Lambda
}\big( e^{-\beta \tau }g \big)^{\Lambda } \phi _{0}\phibar _{x} ,
\end{equation}
where $\mu _{\Lambda }$ is the Gaussian form $\exp (-S_{A} )$ with $A$
equal to the inverse of $U^{\Lambda }$ with  $\beta =0$.

\emph{Scaling:}  This is a transformation $x \mapsto L^{-1}x$ that
maps the hierarchical lattice to itself  by identifying all
points that lie in the same ball of diameter $L$ in the hierarchical
lattice so that they become a single point in a new hierarchical
lattice.  Thus it is the canonical projection (of groups), $\cG
_{N}\rightarrow \cG _{N}/\cG _{1}$, rewritten as
\[
 L^{-1}x=(\dots,x_3,x_2,x_1),\qquad\mbox{for }\,
 x=(\dots,x_3,x_2,x_1,x_0),
\]
which maps $\Lambda = \cG _{N}$ to $ \Lambda/L := \cG
_{N-1}$. Associated to these lattices we have manifolds $\CC ^{\Lambda
}$ and $\CC ^{\Lambda /L}$. A point in $\CC ^{\Lambda }$ is specified
by $(\phi _{x})_{x\in \Lambda }$. Scaling therefore maps a point $\phi
$ in $\CC ^{\Lambda /L}$ \emph{backward} to a point $\cS \phi$ in $\CC
^{\Lambda }$ according to
\[
\cS \phi _{x} \equiv (\cS \phi )_{x}= L^{-1}\phi _{L^{-1}x},
\]
so $\cS \phi $ is constant on cosets $x + \cG _{1}$. The prefactor
$L^{-1}$ is put there to make the RG map, to be defined below,
autonomous.  Functions and forms on $\CC^{\Lambda}$  are
mapped \emph{forward}. For example, for $x \in \Lambda $, $d\phi _{x}$
is a form on $\CC ^{\Lambda }$.  Under scaling we get
\[
\cS ( d\phi _{x}) = L^{-1}d\phi _{x/L} ,
\]
which is a form on  $\CC ^{\Lambda /L}$. We define scaling on
covariances by 
\[
\cS C^{\Lambda /L} (x,y) \equiv (\cS C^{\Lambda /L}) (x,y) = L^{-2}C
(L^{-1}x,L^{-1}y).
\]
Covariances are functions on the lattices so they are mapped
\emph{backwards}.  To summarize: the direction of maps may appear
reversed, but observe that the projection $\Lambda \rightarrow \Lambda
/L$ induces a map backwards of manifolds $\CC ^{\Lambda /L}
\rightarrow \CC ^{\Lambda }$ because $\CC ^{X} = \text{ map } (X
\rightarrow \CC )$ and so forms/functions on the manifold $\CC
^{\Lambda}$ are mapped { forward} to forms/functions on the manifold
$\CC ^{\Lambda /L}$.

The renormalization group rests on the following \emph{scaling
decomposition} 
\begin{equation}\label{eq:dirichlet-scale-decomp}
U^{\Lambda } (\beta ,x) = \cS U^{\Lambda /L} (L^{2}\beta ,x) +
\Gamma(\beta ,x) .
\end{equation}
 The important properties of $\Gamma$ defined by this formula are
that $\Gamma$ is positive semi-definite and finite range.  In
Appendix~\ref{appendix:dirichlet-bc} we prove that 
\begin{equation}\label{eq:dirichlet-scale-decomp-b}
\Gamma (\beta,x) = \frac{1}{1 +\beta }
(\one _{\cG _{0}} (x) - L^{-4} \one_{\cG _{1}}(x)).
\end{equation}
 $\Gamma$ also has the inessential properties $\sum_{y}\Gamma
(\beta ,y)=1$ and $\Gamma (\beta ,y) = \Gamma (\beta ,y')$ for
$y,y'\not =0$ which lead to simplifications specific to this model,
notably Lemma~\ref{lem:no-field-strength}. 

Let $C_{xy}$ with $x,y \in \Lambda $ be an invertible matrix and let
$A$ be the inverse of $C_{xy}$. Define $\mu _{C} = \exp (-S_{A} )$.
According to Lemma~\ref{lem:gaussian1}, $\int \mu _{C}\phi _{a}\phibar
_{b} = C_{ab}$ whenever the inverse $A$ has positive-definite real
part. The $\Gamma $ appearing in (\ref{eq:dirichlet-scale-decomp}) is
only positive semi-definite because, as an operator on $\CC ^{\Lambda
}$, it has the kernel $\Xi ^{\perp}$ consisting of all $\phi _{x}$
that are constant on cosets $x + \cG _{1}$, because $\sum
_{y}\Gamma_{x-y}=0$.  Let $\Xi \subset \CC ^{\Lambda }$ be the
subspace orthogonal to the kernel of $\Gamma $. $\Gamma $ restricts to
the invariant subspace $\Xi$.  The restricted $\Gamma $ is
positive-definite and invertible. We choose coordinates $\zeta $ in
$\Xi $ by picking any basis and define $\mu _{\Gamma}$ as a form on
$\Xi$ using the inverse of $\Gamma$ computed in this basis. The form
$\mu _{\Gamma }$ is independent of this choice of basis because forms
are coordinate invariant. The matrix $\cS U^{\Lambda /L}$ also has no
inverse, but by the same reasoning defines a form $\mu _{\cS
U^{\Lambda /L}}$ on $\Xi ^{\perp}$.  The action of $\cS $ on
covariances was defined so that
\begin{equation}\label{eq:scale1}
\int \mu _{\cS U} \, u = \int \mu _{U}  \,  \cS u ,
\end{equation}
where $U = U^{\Lambda /L}$ and $u$ is a form on on $\Xi
^{\perp}\approx \CC ^{\Lambda /L}$.

Let $g (\phi ) = \sum g_{\alpha ,\bar{\alpha}} (\phi ) d\phi ^{\alpha
}d\phibar ^{\bar{\alpha }}$ be a form on $\CC ^{\Lambda }$. Then
$g(\phi +\zeta)$ is a form on $\Xi ^{\perp} \times \Xi $ defined by
pullback, \ie substituting $\phi +\zeta $ for $\phi $.  Define the
form $\mu _{\Gamma}*g$ on $\Xi ^{\perp}$ by
\[
\mu _{\Gamma}*g (\phi) =  
\int_{\Xi} \mu _{\Gamma } (\zeta) g (\phi +\zeta ).
\]
The scale decomposition (\ref{eq:dirichlet-scale-decomp}) leads to the
\emph{convolution property}
\begin{equation}\label{eq:convolution}
\int _{\CC ^{\Lambda }} \mu _{U} (\phi) F(\phi) = \int _{\Xi ^{\perp}}
\mu _{\cS U} (\phi) \mu _{\Gamma } \ast F(\phi) ,
\end{equation}
which is valid for $F(\phi) $ any smooth bounded form.  This claim
follows by changing variables $(u,v) = (\phi +\zeta , \zeta )$ and
integrating out $v$ using
Corollary~\ref{cor:gaussian-form-convolution}.

\begin{definition} Define the linear operator $\T$ that maps forms on
$\CC ^{\Lambda }$ to forms on $\CC ^{\Lambda /L}$ by $\T u = \cS \mu
_{\Gamma (\beta )} \ast u$. $\T$ is called a \emph{renormalization
group} (RG) transformation.
\end{definition}

\begin{proposition} \label{prop:RG1} 
\[
\int_{\CC ^{\Lambda}} \mu _{\Lambda } \, 
e^{-\beta  \int \tau }u 
= 
\int_{\CC ^{\Lambda/L}} \mu _{\Lambda /L}\, 
e^{- (L^{2}\beta) \int \tau }\T u
\]
\end{proposition}

\proof{} By the $\tau $ isomorphism, the covariance of the Gaussian
$\mu _{\Lambda } \, \exp (-\beta \int \tau )$ is the same as the
covariance of $\mu _{U}$ when $U$ has parameters $\Lambda , \beta
$. Therefore the two Gaussian forms are equal.  By the convolution
property (\ref{eq:convolution}), $\int \mu _{U} u = \int \mu _{\cS U}
\, \mu _{\Gamma } \ast u$ where $U$ in $\cS U$ has scaled parameters
$\Lambda /L$ and $L^{2}\beta $. Apply (\ref{eq:scale1}). \qed

\begin{lemma}\label{lem:commutes-with-Q} $\T $ commutes with the
supersymmetry operator $Q$ defined in (\ref{eq:Q}). Furthermore, if
$u$ is an even supersymmetric form on $\CC ^{x+\cG _{1}}$, then there
is a unique function $f$ such that $\T u = f (\tau _{x})$.
\end{lemma}

\proof{}
$Q$ commutes with $\cS $ because $\cS $ is a pullback. By
Lemma~\ref{lem:Q-commutes-with-int}, $Q$ also commutes with
integrating out.  Thus $[\T ,Q]=0$ because $\mu _{\Gamma }$ is
supersymmetric. The existence of $f$ follows from
Lemma~\ref{lem:supersymmetric-functions}. \qed

Let $X$ be a subset of $\Lambda $.  A form $F_{X}$ is said to be
\emph{localized} in $X$ if it is a form on $\CC ^{X}$. Since Gaussian
random variables are independent if their covariance vanishes, we have
the \emph{independence property}
\[
\mu _{\Gamma }\ast (F_{X} G_{Y}) = 
(\mu _{\Gamma }\ast F_{X}) \ (\mu _{\Gamma }\ast G_{Y}),
\]
whenever the hierarchical distance between $X$ and $Y$ exceeds the
range of $\Gamma $. Given forms $g_{x}$ localized at single sites
$\{x\}$ let
\[
	g^{X}
=
	\prod _{x\in X}g_{x}.
\]
Then the independence property implies
\begin{equation}\label{eq:independence}
\T g^{\Lambda } = 
\prod _{x \in \Lambda/L} \T g^{x+\cG _{1}}.
\end{equation}
$\T g^{x+\cG _{1}}$ is a form on $\CC ^{\{x \}}$.  By
Lemma~\ref{lem:commutes-with-Q}, it has the form $g_{\text{new}} (\tau
_{x})$ for some function $g_{\text{new}} (t)$ on $\RR $. This is a
marvelous property of the hierarchical lattice because it means that
the RG map preserves the multiplicativity of the interaction:
\[
\T g^{\Lambda } = g_{\text{new}}^{\Lambda /L} ,
\]
and therefore $\T$ can be described by the map $g \rightarrow
g_{\text{new}}$.   This has come about because the hierarchical
topology has no overlapping neighborhoods: any pair of neighborhoods
are either disjoint or nested.  

However, there is some redundancy in the pair $(\beta ,g)$: as noted in
(\ref{eq:shift-by-kappa}), the Green's function $G_{g} (\beta ,x)$
depends only on the combination $g (t)\exp (-\beta t )$.  We will
remove this redundancy by imposing the normalization condition
\begin{equation}\label{eq:normalization1}
g  (t) = 1 + O (t^{2}) \text{ as } t \rightarrow 0.
\end{equation}
This normalization assumes that $g _{\text{new}}(0) =1$.  This is true
for the initial interaction $g (t) = \exp (-\lambda t^{2} )$ and, by
Lemma~\ref{lem:gaussian2}, it also holds for $g_{\text{new}}$.  After
each map by $\T$, which, referring to Proposition~\ref{prop:RG1},
takes $(\beta ,g)$ to $(L^{2} \beta ,g_{\text{new}})$, the pair
$(L^{2} \beta ,g_{\text{new}})$ is replaced by an equivalent pair
$(\beta ',g') = (L^{2} \beta +\nu ' ,\exp (\nu ' t )g_{\text{new}})$
to restore (\ref{eq:normalization1}). 
 
\begin{definition}\label{def-T} Let $g$ be a smooth bounded function
on $\RR $ and let $g_{x} = g (\tau _{x})$.  $T_{\beta }g$ is the
function on $\RR $, given by
\[
T_{\beta }g (\tau _{x}) = e^{\nu ' t} \T g^{x+\cG _{1}} ,
\]
where $\nu ' $ is chosen so that $T_{\beta }g (t) = 1 + O (t^{2})$.
\end{definition}

In (\ref{eq:tau-with-g}) there is also the factor $\phi _{0}\phibar
_{x}$. By the independence property (\ref{eq:independence}), the RG
acts on each of the factors $\phi _{0}$ and $\phi _{x}$ independently
if $|x|>L$.   In fact, in this model, the RG acts simply by scaling:

\begin{lemma} \label{lem:no-field-strength} Suppose that $g = a
(\phi ) + b (\phi ) \, d\phi \, d\phibar $ is an even form on $\CC
$. Define $g_{x}$ by replacing $\phi $ by $\phi _{x}$, then
\[
\T g^{x+\cG _{1}} \phi _{x} =
(\T g^{x+\cG _{1}}) \, \cS \phi _{x}.
\]
\end{lemma}
\proof{} Let $G=g^{x+\cG_1}$ with $L^{-2}\phi _{x/L}+\zeta _{x}$
substituted in place of $\phi _{x}$, then
\[
\T g^{x+\cG _{1}} \phi _{x} =
\left(\cS \int \, \mu _{\Gamma } \, G\right) \, \cS \phi _{x} + 
\cS \int \, \mu _{\Gamma } \, G \, \zeta_{x} ,
\]
so we have to prove that $\int \mu _{\Gamma } G \zeta_{x}=0$.  By $\cG
_{1}$ invariance, $\int \mu _{\Gamma } G \zeta_{x}=\int \mu _{\Gamma }
G \zeta_{y}$ for all $y \in x +\cG_1$. Since $\zeta \in \Xi$, $\sum
_{y\in \cG }\zeta _{y}=0$. \qed

By (\ref{eq:tau-with-g}), Proposition~\ref{prop:RG1} and
Lemma~\ref{lem:no-field-strength} we have proved that
\begin{proposition}\label{prop:RG-on-Green} For $|x|>L$,
\begin{equation}\label{eq:T}
G_{g}^{\Lambda } (\beta ,x) = \cS G_{g'}^{\Lambda /L} (\beta ',x) ,
\end{equation}
where $g' = T_{\beta }g$ and $\beta ' = L^{2}\beta  + \nu ' $ with
$\nu ' $ as in Definition~\ref{def-T}.
\end{proposition}

What happens if $|x| \le L$?  The transformation of the factor
$\phi _{0}\phibar _{x}$ is no longer simple, but the next result says
that it reproduces itself together with supersymmetric corrections.
\begin{lemma}\label{lem:observable-form}
If $u$ is a smooth supersymmetric even form on $\CC ^{\cG _{1}}$, then
there are unique functions $f_{1},f_{2}$ such that $\T u \phi_{0}
\phibar _{x} = f_{1} (\tau _{0})+ f_{2} (\tau _{0})\phi_{0} \phibar
_{0}$.
\end{lemma}

\proof{} Let $v= \T u \phi _{0}\phibar _{x}$ and $w=Qv$, then $w$ is a
supersymmetric form of odd degree. By
Lemma~\ref{lem:supersymmetric-functions}, $w = a (\tau ) (\phi
d\phibar + \phibar d\phi)$.  The solutions of $w=Qv$ are $v = - (2\pi
i)^{-1} a (\tau )\phi \phibar + b (\tau )$. $a$ and $b$ are unique
because the degree zero part of $v$ determines the combination $-
(2\pi i)^{-1}a (t)t + b (t)$ and the degree two part determines $-
(2\pi i)^{-1}a' (t)t + b' (t)$.\qed

\section{Coordinates for Interactions}\label{section:coordinates}
\setcounter{theorem}{0}
\setcounter{equation}{0}

\newcommand{\vt}{\tilde{v}}
\newcommand{\lambdat}{\tilde{\lambda }}
\newcommand{\nut}{\tilde{\nu }}
\newcommand{\rmain}{r_{\mathrm{main}}}
\newcommand{\obs}{\mathcal{O}} 
\newcommand{\obst}{\tilde{\obs}}

The parameters in the Green's function (\ref{eq:rg-trans1}) are
$\beta$, a smooth function $g$, and the volume $\Lambda $.  $g$ defines
a coupling constant $\lambda $ and a smooth function $r (t)$ by
\[
g (t ) = e^{-\lambda t ^{2}} + r (t ) \text{ with }
r (t) = O (t^{3}) \text{ as } t \rightarrow 0,
\]
because by Definition~\ref{def-T}, $\beta $ is adjusted so that $g
(t) = 1 + O (t^{2})$. $r$ will be called the \emph{remainder}.

Consequently, we may describe the map $g \rightarrow T_{\beta }g$ by
its action on the parameters $ \beta ,\lambda ,r \rightarrow \beta ', \lambda ', r '$
, where $\beta '$ , $\lambda '$ and $r '$ solve
\begin{align}\label{eq:coordinates2}
&
e^{-L^{2}\beta \tau }\T (e^{-\lambda \tau  ^{2}} +r)^{\cG _{1}} = 
e^{- \beta ' \tau  } \left(
e^{-
\lambda ' \tau ^{2} } + r '
\right), 
\ \ \
r ' (t) = O (t^{3}).
\end{align}
Iteration of this map defines a finite sequence $(\beta _{j}, \lambda
_{j}, r_{j},\Lambda _{j})_{j=0,\dots ,N-1}$. The sequence terminates
because the initial $\Lambda _{0} =\cG _{N}$ is scaled down by $L$
with each RG map and eventually becomes $\cG _{1}$.  Then there is one
final integration.  This sequence exists for any initial choice of
parameters with $\Re \lambda _{0} >0$ because $\Lambda _{0}$ is
finite.  

This hierarchical model has the nice feature that enlarging the
initial volume $\Lambda _{0}$ merely extends the sequence --- the
longer sequence coincides with the shorter for shared indices
$j$.  Therefore the sequences consistently extend to an infinite
sequence.

In (\ref{eq:tau-with-g}) we rewrite the \emph{observables} $\phi _{0}$ 
and $\phi _{x}$ as if they were part of the interaction by
\[
g^{\Lambda }\phi _{0}\phibar _{x} = \frac{d}{d\gamma }_{|_{0}}
g^{\Lambda } e^{\gamma \obs} ,
\]
with $\obs = b_{1}\phi _{0}\phibar _{x}$ and $b_{1}=1$. Then
Lemma~\ref{lem:no-field-strength} asserts that $\T$ acts on $\gamma \obs$
at order $\gamma $ by $x \rightarrow L^{-1}x$ and $b_{1}\rightarrow
L^{-2}b_{1}$ to produce iterates $x_{j}:=L^{-j}x$ and
$b_{i,j}:=L^{-2j}b_{1}$ for $j=1,\dots ,N (x)-1$, such that
$|x_{j}|\ge L$.  $N (x)$ is the number of iterations, $\log _{L}|x|$,
that are needed to scale $x$ to $0$. For $j\ge N (x)-1$,
Lemma~\ref{lem:observable-form} asserts that $\mathrm{T}_{j}:=
\mathrm{T}_{\beta _{j}}$ maps $\gamma \obs_{j}$ into
\[
\obs_{j+1} = f_{1,j+1} (\tau _{0}) + f_{2,j+1} (\tau _{0})\phi _{0}\phibar
_{0}. 
\]
 In this and subsequent calculations functions of $\gamma$ are
identified with their linearizations because we only need
to know the derivative with respect to $\gamma$ at $\gamma =0$. 
The Green's function can be accurately calculated without complete
knowledge of $f_{1,j}$ and $f_{2,j}$. To this end we define
\begin{equation}\label{eq:defines-v}
v_{x} = 
\begin{cases}
\lambda \tau _{x}^{2} &	\text{ if } x \not =0\\
\lambda \tau _{0}^{2} - \gamma (b_{0}+b_{1}\phi _{0}\phibar _{0} +
b_{2}\tau _{0}\phi _{0}\phibar _{0} + b_{3}\tau _{0}) & \text{ if } x =0
\end{cases}
\end{equation}
and consider the action of $\T$ on $g^{\Lambda }$ when $g_{x} =
e^{-v_{x}} + r_{x}$.  Part of the observable $\obs$ is in $v$, and the
rest of it, which is the part we will not need to calculate in detail,
is in $r$.  The split is uniquely determined by

\begin{definition} \label{def:normalization} Let $r$ be a form
localized at a single lattice site.  We say $r$ is \emph{normalized}
if $\frac{d^{q}}{dt^{q}}r_{t}=0$ at $t=0$ for $q=0,1,\dots ,5$, where
$r_{t}$ is defined by replacing $\phi $ by $t\phi$ in $r$ including in
$d\phi $ and $d\phibar $.

The interaction $v$ is said to be normalized if $v = \lambda \tau ^{2}
+ \gamma \obs $ for some $\lambda $ and $\obs$ is an even polynomial
form of degree less than or equal to four.
\end{definition}

The action of $\T$ on $g^{\Lambda }$ is now completely described by
the action on parameters
\[
(\beta ,\lambda ,b,r,\Lambda ) \rightarrow (\beta ',\lambda
',b',r',\Lambda ') ,
\]
where $b = (b_{0},b_{1},b_{2},b_{3})$.  The Green's function is 
\begin{gather}
G_{\lambda _{0}}^{\Lambda _{0}} (\beta _{0},x) 
= \frac{d}{d\gamma }_{|_{0}} 
\int \mu _{\Lambda _{0}} 
\left( 
e^{-\beta _{0}\tau -\lambda
_{0}\tau ^{2}}
\right)^{\Lambda _{0}}
e^{\gamma \phi _{0}\phibar _{x}}\nonumber \\
= \frac{d}{d\gamma }_{|_{0}} \int \mu _{\cG _{1}} 
\left(
e^{-\beta _{N-1}\tau}[e^{ -v_{N-1}}+r_{N-1}]
\right)^{\cG _{1}}\nonumber \\
=: b_{0,N} \label{eq:green-and-bzero} ,
\end{gather}
because the final integration with respect to $\mu _{\cG _{1}}$ is
being considered to be a final RG map\footnote{with $\Gamma $ replaced
by $U^{\cG _{1}}$} followed by setting $\phi $ and $d\phi $ to zero so
that only the $b_{0}$ part of $v$ ends up in the final result.

Note that the normalization condition is designed so that there is no
$\gamma $ dependence in the sequence $(\beta _{j}, \lambda _{j})$.

A surprising fact is that the $b_{3}\tau $ term never plays any role
beyond being there!  No term of the form $\gamma F (\tau )$ with $F
(0) = 0$ contributes to the Green's function because $( d/d\gamma
)_{0} \int \, \mu \, G (\tau ) \, \exp (\gamma F (\tau ) ) = G (0) F
(0)$ by Lemma~\ref{lem:gaussian2}.  For this reason, we leave out the
$b_{3}$ terms in the rest of this paper.

\section{Second Order Perturbation Theory}
\label{section:second-order-perturbation-theoryI}
\setcounter{theorem}{0}
\setcounter{equation}{0}

\newcommand{\Vt}{\tilde{V}}
\newcommand{\bt}{\tilde{b}}
\newcommand{\betat}{\tilde{\beta }}

We have shown that the RG induces a map from parameters $(\beta
,\lambda ,b,r)$ to $(\beta ',\lambda ',b',r')$. We will also write $v
\rightarrow v'$ recalling that $v$ is determined by
(\ref{eq:defines-v}). In this section we construct an approximation 
\[
v \rightarrow \vt , \hspace{2mm} (\beta ,\lambda ,b) \rightarrow
(\betat ,\lambdat ,\bt )
\]
to the exact map using second order perturbation theory in powers of
$\lambda $.  This approximation plays a major role in determining the
$\log ^{\frac{1}{8}}$ corrections in the end-to-end distance of the
interacting walk.

\emph{Notation: } 
\begin{equation}\label{eq:notation-perturbation-theory}
B = 1- L^{-4}, \hspace{2mm}
B_{p} = \int \Gamma ^{p} (y)\, dy.
\end{equation}
$B_{p}$ is a function of $\beta $ through $\Gamma $. In particular, 
\begin{equation}\label{eq:notation-perturbation-theory2}
B_{1} = 0, \hspace{2mm} B_{2} = B (1+\beta )^{-2}, \hspace{2mm}
\Gamma (0) = B (1+\beta )^{-1}.
\end{equation}

We will show that the second order approximation to $(\beta ,\lambda ) 
\rightarrow (\beta ',\lambda ') $ is
\begin{align}\label{eq:recursion-beta-lambda}
	\betat 
&=
	L^{2} (\beta  + 2 \Gamma (0)\lambda + O(\Gamma ^{3})\lambda ^{2})
	\nonumber \\[2mm]
 	\lambdat
&=
	\lambda - 8 B_{2} \lambda ^{2} ,
\end{align}
where $ O(\Gamma ^{p})$ is an analytic function\footnote{They are
integrals of $p$ or more covariances, $\Gamma (\beta) $ and polynomial
in $\lambda $. They can be computed explicitly using Feynman diagrams
as in Appendix~\ref{appendix:feynman-diagrams}. } of $\beta $ and
$\lambda $ that is bounded in absolute value by $c |1+\beta |^{-p}$,
for $|\lambda |$ bounded.

Likewise, the parameters $b$ have the approximate recursion
\begin{align}\label{eq:recursion-b}
        \bt _{0,j+1} 
&=
        b_{0,j} + \Gamma _{j }(x_{j}) b_{1,j} + O(\Gamma _{j}^{3}) \lambda _{j} b_{1,j} + 
        O (\Gamma _{j}^{2}) b_{2,j}
	\nonumber \\[2mm]
	\bt _{1,j+1} 
&= 
        L^{-2} \left[b_{1,j} + O (\Gamma _{j}^{2}) \lambda _{j} b_{1,j} + 
        O (\Gamma _{j}) b_{2,j} \right]
	\nonumber \\
        \bt _{2,j+1} 
&=
        L^{-4} \left[
	b_{2,j} + O (\Gamma _{j}^{2})\lambda _{j} b_{2,j} 
	\right] ,
\end{align}
where the $j$ subscript on $\Gamma $ means that $\beta =\beta _{j}$. 
All the terms involving $\Gamma $ vanish for $j<N (x)-1$.  See
Section~\ref{section:coordinates}.

\begin{proposition}\label{prop:recursion1} Let $\rmain = \T (
e^{-v})^{\cG _{1}} - e^{ - \vt -\nut \tau }$ with $\nut = \betat
-L^{2}\beta $. Then $\rmain = O (\lambda ^{3}) + O(\lambda ^{2}\gamma
)$ as a formal series in powers of $\lambda$.
\end{proposition}

To prove Proposition~\ref{prop:recursion1} we introduce the following
Laplacian
\begin{equation}\label{eq:def-laplacian}
\Delta _{\Gamma} = \sum_{x,y} \Gamma (x-y) \bigg (
	\frac{\partial }{\partial \phi_{x} } 
	\frac{\partial }{\partial \phibar _{y} } +
	\frac{\partial }{\partial \psi_{x} } 
	\frac{\partial }{\partial \bar{\psi}_{y} } 
	\bigg).
\end{equation}
The partial derivatives $\partial/\partial \psi_{x}$ are formal
anti-derivatives (equivalently, interior products with vector
fields dual to the forms $\psi _{x}$). Let $F$ be a smooth bounded
form, then
\begin{equation}\label{eq:laplacian-and-mu}
	\mu _{t\Gamma }\ast F 
=
	F + t \Delta _{\Gamma }F + O (t^{2})
	\text{ as } t \rightarrow 0.
\end{equation}
Given forms $X, Y$ we define a new form $X\join_{\Gamma } Y$ by
\[
	\Delta _{\Gamma }XY 
= 
	(\Delta _{\Gamma }X)Y + X\Delta _{\Gamma }Y +
	X\join_{\Gamma } Y.
\]
Therefore, denoting partial derivatives with respect to $\phi _{x}$
and $\phibar _{x}$  by subscripts,
\[
	X\join_{\Gamma } Y
=
	\sum_{x,y} \Gamma (x-y) \bigg (
	X_{\phi _{x}} Y_{\phibar _{y}} +
	X_{\phibar _{x}} Y_{\phi_{y}} 
	+ 
	(-1)^{\text{sgn} (X)}
	X_{\psi _{x}} Y_{\bar{\psi} _{y}} -
	(-1)^{\text{sgn} (X)}
	X_{\bar{\psi} _{x}} Y_{\psi _{y}}
	\bigg).
\]
Thus we can define $X \join _{\Gamma }\join _{\Gamma } Y$ by applying
the second $\join _{\Gamma } $ to each term on the right-hand side of
this equation.

A \emph{polynomial} form is a form whose coefficients are polynomials
in $\phi$ and $\phibar $.  The essential property of such forms is
that they are annihilated by $\Delta _{\Gamma }^{j}$ for $j>>1$.

Let $V$ be any polynomial form and set
\[
V_{t} = e^{t\Delta _{\Gamma }}V,
 \hspace{3mm}
\cL = \frac{\partial }{\partial t} - \Delta _{\Gamma}.
\]
Then $V_{t}$ has the important property $\cL V_{t} = 0$. 

\begin{lemma}\label{lem:Q} For any polynomial form $V$,
\begin{eqnarray*}
	Q_{t} 
&= 
	\frac{1}{2} \sum _{j\ge 1}\frac{1}{j!}
	V_{t}\join _{t\Gamma }^{j} V_{t}
\end{eqnarray*}
satisfies
\[
	\cL [-V_{t} + Q_{t}]
=
	\frac{1}{2} V_{t} \join_{\Gamma } V_{t}.
\]
\end{lemma}

\proof{} Since $V$ is a polynomial form, the sum over $j$ terminates
after finitely many terms, so $Q$ is defined.  Furthermore $\cL
V_{t}=0$. Therefore,
\begin{align*}
	\cL \bigg (V_{t}\join _{t\Gamma } V_{t} \bigg)
&=
        V_{t}\join _{\Gamma } V_{t} -  
	V_{t}
        \join _{\Gamma } \join _{t\Gamma }
        V_{t} 
\\
	\cL \bigg (\frac{1}{2!}V_{t}\join _{t\Gamma }^{2} V_{t} \bigg)
&=
        V_{t}
        \join _{\Gamma }\join _{t\Gamma }
        V_{t} -
        \frac{1}{2!}V_{t}
        \join _{\Gamma }\join _{t\Gamma }^{2}
        V_{t} ,
\end{align*}
together with similar equations for the remaining terms in
$Q_{t}$. Add these equations. \qed

\newcommand{\Vhat}{\hat{V}}
\begin{lemma}\label{lem:second-order-perturbation} Let $\Vhat_{t} = [V_{t}
- Q_{t}]_{t=1}$.  Then $\mu _{\Gamma } \ast e^{-V} - e^{-\Vhat_{t} } = O
(V^{3})$ where $O (V^{3})$ means that the formal power series in
powers of $\alpha $ obtained by replacing $V$ by $\alpha V$ in the
left-hand side is $O (\alpha ^{3})$.
\end{lemma}
We call $\Vhat_{1} $ the \emph{second order perturbative effective
interaction}.

\proof{} We use the \emph{Duhamel formula:} 
Let $W_{t}$ be any smooth family of forms with $W_{0}=\exp (-V
)$. Then, using (\ref{eq:laplacian-and-mu}),
\begin{align*}
	\mu _{\Gamma } \ast e^{-V} - W_{1}
&=
	\int _{0}^{1} \,dt \, \frac{d}{dt} \mu _{(1-t)\Gamma } 
	\ast W_{t}\\
&=
	\int _{0}^{1} \,dt \, \mu _{(1-t)\Gamma } 
	\ast \cL W_{t}.
\end{align*}
Choose $W_{t} = \exp (-\Vhat_{t})$.  By Lemma~\ref{lem:Q} $\cL W_{t} = O
(\lambda ^{3})$ because
\begin{align}\label{eq:pert1}
\cL e^{\Vhat_{t}}
&= e^{\Vhat_{t}}\left(
\frac{1}{2}V_{t}\join _{\Gamma }V_{t} -
\frac{1}{2}\Vhat_{t}\join _{\Gamma }\Vhat_{t} \right)\nonumber\\
&= e^{\Vhat_{t}}\left(
Q_{t}\join _{\Gamma }V_{t} -
\frac{1}{2}Q_{t}\join _{\Gamma }Q_{t} \right) = O(V ^{3}).
\end{align}
\qed 

\Proof{ of Proposition~\ref{prop:recursion1}} 
By Lemma~\ref{lem:second-order-perturbation} applied to $V:=\int
_{x+\cG _{1}} v_{y} \,dy$, with $v_{y}$ as defined in
(\ref{eq:defines-v}) 
\[
e^{-L^{2}\beta \tau _{x}}\T e^{-V} 
= e^{-L^{2}\beta \tau _{x}}\cS e^{-\Vhat _{1}}
= e^{-L^{2}\beta \tau _{x}} e^{-\cS\Vhat _{1}}.
\]
$\cS\Vhat _{1}$ is a priori a polynomial of degree six, but in fact
the degree six part vanishes because it contains, from $\join $, a
factor $B_{1} = 0$.  Therefore, $\cS\Vhat _{1}$ is a polynomial form
of degree four. It contains a part $\nut \tau _{x}$ which is absorbed
into $\betat $, so that $\betat = L^{2}\beta +\nut $ and the remaining
part $\cS\Vhat _{1} -\nut \tau _{x} =: \vt _{x}$ is in the form
(\ref{eq:defines-v}) with coefficients $(\lambdat , \bt )$. 

{ Details
for deriving the formulas (\ref{eq:recursion-beta-lambda},
\ref{eq:recursion-b}) for $(\lambdat , \bt )$ are in
Appendix~\ref{appendix:feynman-diagrams}. The main points} 
are as follows: Suppose
coefficients $b, \lambda $ are assigned minus the degree of the
monomial they preface.  Thus $b_{0}$ has degree zero, $b_{1}, 
\nu $ have degree $-2$ and $b_{2}, \lambda$ have degree $-4$.  Let
$\Gamma$ have degree $2$.  Then, for example, a term such as
$b_{2}\lambda O (\Gamma^{2})$ can appear in the right-hand side of the
$b_{2}$ recursion because it has degree $-2$ which equals the degree
of $b_{2}$.  There are certain terms that do not appear because they
contain a factor $\Gamma$ in the form $B_{1} = \sum \Gamma (y) =0$.
In fact, for this reason the $b$ recursion is triangular. \qed

\emph{The Large Field Problem: } We are confining ourselves at present
to formal power series statements because $W_{t}=\exp (-\Vhat _{t} )$
is not integrable: unlike $\cS \Vhat $, the sixth degree part of
$\Vhat $ does not vanish and $W_{t}$ consequently fails to be
integrable.   When we prove estimates on remainders in
Proposition~\ref{lem:r-main-estimates}, we will use another choice of 
$W_{t}$ which is the same up to order $O (V^{3})$.


\section{The Green's function} \label{section:greens-function}
\setcounter{theorem}{0}
\setcounter{equation}{0}

In this section we will prove the main result in this paper,
Theorem~\ref{theorem:main}. 

Recall that in paper I we introduced the enlarged domains
\begin{align*}
&\Dbarb = \{\beta \neq 0: \ |\arg
\beta| < b_\beta + \frac{1}{4} b_\lambda +\epsilon\}; \nonumber \\[4mm] 
&\Dbarb (\rho) =
\Dbarb + \cB(\rho) \text{ with }
\cB(\rho) = \{\beta: \
|\beta| < \rho\}; \nonumber \\[4mm]
&\Dbarlambda = \{\lambda: \ 0 < |\lambda| <
\overline{\delta} \mbox{ and } |\arg \lambda| <
b_\lambda+\epsilon\}.
\end{align*}
We will need

\begin{proposition}\label{prop:beta-lambda-recursion} Let $(\beta
_{0},\lambda _{0})$ be in the domain $\Dbarb\left(\frac{1}{2}\right)
\times \Dbarlambda $  with $\bar{\delta}$ sufficiently small.  The
sequence $(\beta_j,\lambda_j)_{j=0,1,\ldots,M}$ is such that 
\begin{eqnarray}\label{eq:1.16}
\lambda_{j+1} & = & \lambda_j - \, \frac{8B
\lambda^2_j}{(1+\beta_j)^2} + \epsilon_{\lambda,j}, \nonumber \\[4mm]
\beta_{j+1} & = & L^2\left[\beta_j + \frac{2B}{1+\beta_j} \,
\lambda_j\right] + \epsilon_{\beta,j},
\end{eqnarray}
where the $\epsilon_{\lambda,j}$, $\epsilon_{\beta,j}$ defined by
these equations are analytic functions of $(\beta_0,\lambda_0)$
satisfying
\begin{eqnarray}\label{eq:1.17}
|\epsilon_{\lambda,j}| & \leq &
c_L|\lambda_j|^3|1+\beta_j|^{-1}, \nonumber \\[4mm]
|\epsilon_{\beta,j}| & \leq &
c_L|\lambda_j|^2|1+\beta_j|^{-2}.
\end{eqnarray}
Here $B = 1-L^{-4}$ and $M$ is the first integer such that
$(\beta_M,\lambda_M)$ is not in the domain
$\Dbarb\left(\frac{1}{2}\right) \times \Dbarlambda $. If no such
integer exists, then $M = \infty$.
\end{proposition}

The formulas (\ref{eq:1.16}) were already obtained in
(\ref{eq:recursion-beta-lambda}).  The new content is the estimate on
the errors.

Recall also that there are observable parameters $b_{j}:= (b_{0,j},
b_{1,j}, b_{2,j})$.  Let $\epsilon _{*,j} = b_{*,j+1} - \bt _{*,j+1}$
where $* = 0,1,2$, be the errors between the exact recursions for
these parameters and the second order perturbative recursions, defined
in (\ref{eq:recursion-b}).  Since we defined the observable to be a
derivative at $\gamma =0$, we may suppose, without loss of generality,
that $\epsilon _{*,j}$ are linear in $b_{j}$. Higher order terms will
drop out when $\gamma $ is set to zero. $O (\Gamma _{j}^{p})$ denotes
an analytic function of $\beta _{0}$ and $\lambda _{0}$ which is 
defined on $\Dbarb\left(\frac{1}{2}\right) \times \Dbarlambda $ and
 bounded in absolute value by $c_{L}|1+\beta _{j}|^{-p}$.

\begin{proposition}\label{prop:error-for-b} For $M \ge j \ge N (x) -1$
 and $q=0,1,2$,
\[
|\epsilon_{q,j}| \le O (\Gamma _{ j }^{3-q}) |\lambda
_{j}|^{2}|b_{j}|,
\]
 where $|b_{j}| := |b_{1,j}| + |b_{2,j}|$.
\end{proposition}
These two theorems will be proved in the next section.  In the
next proof $c_{L}$ denotes constants chosen after $L$ is fixed,
whereas $c$ denotes constants chosen before $L$ is fixed.  The values
of these constants are not relevant to the proof so these symbols can
change values from one appearance to the next. 

\newcommand{\lambdabar}{\bar{\lambda}}
\newcommand{\Gammabar}{\bar{\Gamma}}

\Proof{ of Theorem~\ref{theorem:main}} During this proof we will write
$\lambdabar := \lambda _{N (x)}$, $\Gammabar := \Gamma _{N (x)}$ and
$k:=j - N (x) +1$.  We fix $ \xi  \in (1,2)$. The constant
$\bar{\delta}$ that controls the size of $|\lambda |$ in the domain
$\Dbarlambda $ will be  the minimum of a finite number of choices
that achieve bounds in generic inductive steps. The choices
depend on $L$ and $ \xi  $.  Thus, throughout the proof we
will be using the following principles:
\begin{enumerate}
\item $ \exists c: \hspace{2mm}  |\lambda _{j}| \le c
|\lambdabar |$ for $j\ge N (x)$ by Proposition~I.1.5.
\item $|c_{L}\lambda _{j}|<c$ for any $c$ because the domain for
$\lambda _{0}$ is chosen \emph{after} $L$ is fixed.
\item $1+c_{L}|\lambda _{j} |< \xi  $ because the domain for
$\lambda _{0}$ is chosen \emph{after} $ \xi  $ is fixed.
\item $O (\Gamma _{j})\le c_{L}$ because $\beta _{j} \in \bar{\cD}
_{\beta } (\frac{1}{2})$.
\item For $j\ge N (x)$, $O (\Gamma _{j}) \le c O (\Gammabar )$ because it
holds when $|\beta _{N (x)}| \le 1$ since $\beta _{j} \in \Dbarb
(1/2)$ and it also holds when $|\beta _{N (x)}|>1$ because then the
$\beta $ recursion causes $\beta _{j}$ to grow exponentially.
\end{enumerate}
Recall from Section~\ref{section:coordinates} that for $j\le N (x)-1$,
$b_{0,j}$ and $b_{2,j}$ vanish, while $b_{1,j} = L^{-2j}$.  These
values will start inductive arguments at $j=N (x)-1$. The term
\emph{recursion} denotes the perturbative recursion
(\ref{eq:recursion-b}) combined with the bound on the error,
Proposition~\ref{prop:error-for-b}.

\medskip \noindent \emph{Claim 1:} For $j \ge N (x)$, $|b_{1,j}| \le
 \xi  ^{k}L^{-2j}, \hspace{2mm} |b_{2,j}| \le O (\Gammabar )|\lambdabar
|^{2}  \xi  ^{k}L^{-2j}$.

\medskip \noindent \proof{} By induction using the recursion.

\medskip \noindent \emph{Claim 2:} For $j \ge N (x)$, $|L^{2j}b_{1,j}
-1| \le  \xi  ^{k} O (\Gammabar^2 )|\lambdabar |$.

\medskip \noindent \proof{} Let $r_{k}:= |L^{2j}b_{1,j}-1|$. Then
$r_{0} =0$. By the recursion and claim 1, $r_{k+1} \le r_{k} + O
(\Gammabar )|\lambdabar | \xi  ^{k}$. This implies the claim.

\medskip \noindent \emph{Claim 3:} For $j=N (x)-1$ let $a_j=0$ and
for $j\ge N (x)$ let $a_{j+1} = a_{j} + L^{-2j} \Gamma (\beta
_{j},L^{-2j}x)$. Then
\[
|b_{0,j}-a_{j}| \le L^{-2N (x)} O (\Gammabar ^3)|\lambdabar |.
\] 

\medskip \noindent \proof{} Let $r_{j} = |b_{0,j}-a_{j}|$.  Then
$r_{j}=0$ for $j=N (x)-1$. By the recursion and the previous claims,
\[
r_{j+1} \le r_{j} + L^{-2j}  \xi  ^{k} O (\Gammabar ^3)|\lambdabar |.
\]
This proves the claim.

Recall the definition of $\bhat _{j} = \beta _{j}-\beta ^{c}_{j}$ in
Proposition~I.1.3.  Let $u_{j}:= L^{2[k-1]}\bhat _{N (x)}$. By
Proposition~I.1.5, $u_{j} \in \Dbarb $.

\medskip \noindent \emph{Claim 4:} For $j \ge N (x)-1$,
\[
\left|
\frac{1}{1+\beta _{j}} - \frac{1}{1+u_{j}} 
\right|
=
\left|
\frac{u_{j}-\beta _{j}}{(1+\beta _{j}) (1+u_{j})}
\right|
\le |\lambdabar |  \xi  ^{k} O (\Gammabar ^{2}).
\]

\medskip \noindent \proof{} By Proposition~I.1.3, $(1+\beta
_{j})^{-1}$ is indistinguishable from $(1+\bhat _{j})^{-1}$ up to an
error that can be absorbed in the right-hand side of our claim. By
Lemma~I.4.2, part (4),
\[
\bhat _{j} = u_{j}\prod _{l=N (x)}^{j-1} \left(
1 + O (\lambda _{l}O (\Gamma _{l}))
\right) ,
\]
so
\[
|\bhat _{j} - u_{j}| \le \left|
e^{k |\lambdabar | O (\Gammabar )} -1
\right| |u_{j}|
\le |\lambdabar | O (\Gammabar ) \xi  ^{k}|u_{j}|.
\]
The claim follows because $u/ (1+u)$ is bounded for $u \in \Dbarb $.

\medskip \noindent \emph{Claim 5:} $|a_{N} - G_{0} (\effbeta{N (x)},x)
| \le L^{-2N (x)} | \lambdabar | O (\Gammabar ^{2})$.

\medskip \noindent \proof{} By the definition of $u_{j}$ and
$\effbeta{l} = L^{-2l} \bhat _{l}$,
\[
G_{0} (\effbeta{N (x)},x) = \sum L^{-2l}\Gamma (u_{l},L^{-l}x).
\]
$a_{N}$ is the same expression but with $u_{l}$ replaced by $\beta
_{l}$. Thus the claim is a consequence of claim 4.

\medskip In our first paper we proved that the sequences $(\beta
_{j},\lambda _{j})$ in Theorem~\ref{theorem:main} never exit the
domain $\Dbarb (1/2) \times \Dbarlambda $, so $M = \infty $.  Recall
from (\ref{eq:green-and-bzero}) that the Green's function equals
$b_{0,N}$.  By claims 3 and 5,
\[
|b_{0,N}- G_{0} (\effbeta{N (x)},x)| \le L^{-2N (x)} |\lambdabar | O
(\Gammabar ^{2}).
\]
By Proposition~I.1.4, part (3), the right-hand side is less than
$|G_{0} (\effbeta{N (x)},x)|$. \qed 


\section{Estimates on a Renormalization Group
Step}\label{section:norms}

\setcounter{theorem}{0}
\setcounter{equation}{0}

In this section we will prove
Propositions~\ref{prop:beta-lambda-recursion} and
\ref{prop:error-for-b}.

Let $X$ be a subset of $\Lambda $.  A form $F_{X}$ is said to be
\emph{localized} in $X$ if it is a form on $\CC ^{X}$. Thus $F_{X} =
\sum _{\alpha } F_{X,\alpha } (\phi ) \psi^{\alpha }$ where
$F_{X,\alpha }$ is a smooth function on $\CC ^{X}$ and $ \psi^{\alpha
} = \prod_{x\in X} \psi_{x}^{\alpha _{x}}
\psibar_{x}^{\bar{\alpha}_{x}} $.  In particular, $g^{X}$ defined by
$g^{X}=\prod _{x\in X}g_{x}$ is localized in $X$, when the forms
$g_{x}$ are localized at single sites $\{x\}$.

\newcommand{\fh}{\mathfrak{h}}
\begin{definition} Given $h \ge 0$, $w > 0$, $w^{-X} = \prod _{x\in
X}w ^{-1}(\phi _{x})$, let
\begin{align*}
	\|F_{X}\|_{w,h}
&= 
	\sum _{\alpha ,\beta }
	\frac{h^{\alpha + \beta }}{\alpha ! \beta !}
	\sup_{\phi }|\partial ^{\beta }_{\phi }
	F_{X,\alpha}|w^{-X} ,
\\
	|F_{X} |_{h}
&=
	\sum _{\alpha ,\beta }
	\frac{h ^{\alpha +\beta }}{\alpha ! \beta !}
	|\partial ^{\beta }_{\phi }
	F_{X, \alpha} (0)|.
\end{align*}
\end{definition}

{\noindent These norms measure large- and small-field behaviors, respectively.}

\emph{Complex Covariances and Weights:} The \emph{weight} $w$ is a
positive function used to track decay or growth of the interaction at
$\phi =\infty $. In this paper we use $w (A,\kappa ):= A\exp (-\kappa
|\phi |^{2} )$ with $A\ge 1$ and, for tracking decay, $\kappa $ is
positive.  We want to see how decay is affected by convolution with
Gaussian functions and forms. Let
\begin{equation}\label{eq:rhoC1}
\rho _{\alpha }(\phi ):= \det(\pi \alpha C) ^{-1} e^{-\sum {\alpha ^{-1}\phi _{x}
 C^{-1}_{xy}}\phibar _{y}} ,
\end{equation}
where $\alpha = \exp (i\theta )$ has unit modulus and positive real
part and $C_{xy}$ is a positive-definite matrix with indices with $x,y
\in X$.  By Gaussian integration, $|\rho _{\alpha}|\ast w (A,\kappa
)^{X}$ is bounded by $w (A',\kappa ')^{X}$, where $A' = 2A/\cos \theta
$ and $\kappa '=\kappa /2$, if $\kappa  \|C\|_{\text{operator}}
$ is small enough.  Thus there is a new weight $w_{\alpha C}$ such
that $w_{\alpha C}^{X} \ge |\rho _{C}|\ast w^{X}$.  Then, for any
smooth function $f_{X}$,
\[
|\rho _{\alpha}\ast f_{X}| \le |\rho _{\alpha}|\ast w^{X} \|w^{-X}f_{X}
\|_{\infty }.
\]
By applying this estimate also to derivatives of $f_{X}$, 
\begin{equation}\label{eq:rhoC3}
\|\rho _{\alpha}\ast f_{X} \|_{w_{\alpha C},h} \le \|f_{X}
\|_{w,h}.
\end{equation}

We use the notation $\partial _{\psi }^{\alpha } 
=
	\prod_{x} \partial _{\psi_{x}}^{\alpha _{x}} 
	\bar{\partial }_{\psi _{x}}^{\bar{\alpha }_{x}}
$ where $\partial _{\psi_{x}}$ is a formal anti-derivative with respect to
$\psi _{x}$. 

\begin{lemma} \label{lem:properties-of-norm}  Properties of the norms:
\begin{enumerate}
\item[(i)] $\|g^{X}\|_{w,h} \le \|g \|_{w,h}^{X }$, where
$\|g \|_{w,h}^{X } =
\prod _{x\in X} \|g_{x} \|_{w,h}$  
\item[(ii)] $\|f_{x}g_{x}\|_{w,h} = \|f_{x}\|_{w_{f},h} \,
\|g_{x}\|_{w_{g},h}$ when $w_{f}w_{g} = w$
\item[(iii)] $\| \cS F_{X}\|_{\bar{w},\cS h} \le \|F_{X}\|_{w,h}$
where $\bar{w}\ge \cS w^{|\cG _{1}|}$ and $\cS  h :=Lh$
\item[(iv)] For $0 \le h < h'$, 
$\|\partial_{\psi }^{\alpha } \partial ^{\beta
}_{\phi }F_{X} \|_{w,h} \leq (\alpha +\beta )! (h'-h)^{-\alpha -\beta }
\| F_{X}\|_{w,h'}$
\item[(v)] $
\| \mu _{C}\ast F_{X}\|_{w_{\alpha C},h} \leq 
\exp\left[ \sum_{x,y \in X} | C_{xy} | h^{-2}\right]
\|F_{X}\|_{w,h}$
\end{enumerate}
(i) - (iv) are valid when $\|\cdot \|_{\ast,h}$ is replaced by
$|\cdot |_{h}$.
\end{lemma}

\proof{} Properties (i) -- (iv) need only be proved for the $w,h$ norm
because the $|\cdot |_{h}$ norm is the limit as $\kappa \rightarrow
-\infty $ of the $\|\cdot\|_{w,h}$ norm, when $w = w (1,\kappa )$.

Parts (i) and (ii) are proved on page 103 and Appendix A of
\cite{BEI92}.  (iii) is easy. (iv) is proved on page 104 of
\cite{BEI92}. (v) is also proved starting on page 104, noting that
(\ref{eq:rhoC3}) is (6.4) in \cite{BEI92}.\qed

The results of this section are organized by increasing number of
hypotheses. Each proposition assumes the hypotheses that have been
given earlier.   The important point is that each hypothesis is
satisfied when: (a) $L\ge 2$ is sufficiently large; (b) $|\lambda |$
is sufficiently small \emph{depending on} $L$; and (c) $h = |\lambda
|^{-1/4}$.   

The \emph{constants} $c,c_{L}, c_{q}, \dots $ that appear in
hypotheses and conclusions are numbers in $(0,\infty )$.  $c_{\ast }$
denotes a number in $(0,\infty )$ whose value is permitted to depend
on $\ast $, whereas $c$ is a number determined independently of all
parameters $L,\lambda , \beta $ and others that appear in the
theorems. \emph{These symbols are permitted to change value from one
appearance to the next}.  Constants that always have the same value in
all appearances will be denoted by a letter other than $c$.

\medskip \emph{Hypothesis ($\mu $):} $\mu =\mu _{e^{i\theta }C}$ where
$C_{xy}$ is a positive semi-definite matrix such that $C_{xy} = 0$ if
$|x-y| > L$, $\cos \theta \ge c_{1}$ and $|C_{x,y}| \le c_{2}$. 

The constants $c_{1}, c_{2}$ in this hypothesis can be chosen
arbitrarily. Once they are fixed; estimates below are \emph{uniform} in
the choice of $\mu $. Constants appearing without qualification in
hypotheses can be chosen arbitrarily.

In the following lemma, $\Delta = \Delta _{\Gamma }$ is the Laplacian
defined in (\ref{eq:def-laplacian}) with $\Gamma $ replaced by $\exp
(i\theta )C$ and $\|C \|$ is the maximum of $|C_{xy}|$.  $E_{q}
(\Delta )$ is the power series $\sum _{n\le q}\Delta ^{n}/n!$ for
$\exp (\Delta )$ truncated at order $q$. $X$ is a subset of a $\cG
_{1}$ coset.

\medskip \emph{Hypothesis ($\fh$):} $0 \le \fh \le h$. { The parameter for the small-field norm $|\cdot|_{\fh}$ specifies how large a ``small field'' can be.  In most cases this is determined by the covariance $C$ (so ${\fh} \approx 
\|C\|^{1/2}$) but occasionally it is useful to let it be determined by the interaction (so ${\fh}=h\approx\lambda^{-1/4}$).
}

\begin{lemma} \label{lem:fluct-equals-scaling} ($\cS \mu \approx \cS
$): Let $\bar{w} \ge \cS w_{tC}^{X}$ for $t\in [0,1]$.
\begin{align*}
(i)\hspace{2mm}&
\| 
\cS \mu \ast g^{X } - \cS  E_{q-1} (\Delta ) g^{X }
\|_{\bar{w},\cS h} \le
c_{{ q,}L}h^{-2q} \|C \|^{q} \|g \|_{w,2h}^{X }\\
(ii)\hspace{2mm}&
|\cS \mu \ast
g^{X} - \cS E_{q-1} (\Delta ) g^{X} |_{\fh} \le 
c_{q,L} h^{-2q}\bar{w} (0) \|C \|^{q} \|g \|_{w,h}^{X},
\ \ \ q = 1, 2, \dots \hspace{1mm}.
\end{align*}
\end{lemma}

\medskip
\proof{} Let $\mu _{t}$ be $\mu $ with $e^{i\theta }C$ replaced by
$te^{i\theta }C$. Then
\[
\cS \mu \ast g^{X} - \cS E_{q-1} (\Delta )g^{X} = R_{q},
\text{ with }
R_{q} = \int _{0}^{t} \frac{(1-t)^{q-1}}{(q-1)!}\cS \mu _{t} \ast
\Delta ^{q} g^{X}
\,dt.
\]
By parts (iii) - (v) of Lemma~\ref{lem:properties-of-norm},
\[
\|R_{q} \|_{\bar{w},\cS h} \le c_{q,L} h^{-2q}
 \|C \|^{q} \|g\|_{w,2h}^{X}.
\]
In part (ii), $|R_{q}|_{\fh }$ is bounded by $\bar{w} (0) \|R_{q}
\|_{\bar{w},\cS h/2}$. \qed

\medskip \emph{Hypothesis ($\lambda $):} $\lambda $ lies in
a sector $|\lambda | \le c_{1} \Re \lambda $ in the right half of the
complex plane and $|\lambda | \le c$, where $c$ is small.

The \emph{qualification} ``where $c$ is small'' means that each of the
following results of this section may require a hypothesis $|\lambda |
\le c$, with $c$ determined in the proof.

\medskip \emph{Hypothesis ($h\approx \lambda ^{-1/4}$):} $|\lambda
|h^{4} \in [c^{-1},c]$.  The default choice is $h = |\lambda |^{-1/4}$
but this assumption allows us to assume the same estimates for
\eg $2h$.

\newcommand{\vhat}{\hat{v}} 
\newcommand{\rhat}{\hat{r}}

\medskip \emph{Hypothesis ($\vhat,w$):} $|\nu | \le c_{1} h^{-2} $,
$\vhat = \lambda \tau ^{2} + \nu \tau + \gamma \obs$ and $w (\phi ) =
e^{-a|\phi /h|^{2}}$ with $4c_{1} \le a \le c_{2}$.  The default
choice is $a=1$ and $\nu \le \frac{1}{2}h^{-2}$. The observable $\obs$
is a polynomial as in (\ref{eq:defines-v}) such that $|\gamma b_{1}|h^4 +
|\gamma b_{2}| h^{4}\le c$, where $c$ is small. {\footnote{ When $|x|=L$, the observable is $\gamma\phi_0\bar{\phi}_x$.  But as far as estimates are concerned, the behavior is the same as that of $\frac{\gamma}{2}(\phi_0\bar{\phi}_0 + \phi_x\bar{\phi}_x)$.}}

\medskip 

The hypothesis on the observable ensures that it is small relative to
$\lambda \tau ^{2}$.  In principle $\gamma $ is infinitesimal because
we only use the derivative with respect to $\gamma $ at zero, but it
seems to be simpler to allow a finite variation of $\gamma $. The
$(\vhat ,w)$ hypothesis implies that $|\vhat |_{h} \le c$.

Referring to the paragraph \emph{Complex Covariances and Weights}, we
find that the weight $w$ in the $(\vhat ,w)$ hypothesis is $w
(A,\kappa )$ with $A=1$ and $\kappa =ah^{-2}$. Thus $\kappa \approx
|\lambda |^{1/2}$ is small, as required, and we can choose $\bar{w} =
\cS w (A',\kappa ')^{X}$ in
Lemma~\ref{lem:fluct-equals-scaling}. Next, we can reduce to $A'=1$ in
$w (A',\kappa ')$ by putting a constant $c_{L} = {A'}^{X}$ in front of
the norm. Finally, the rescaling $\cS $ maps $ w (1,\kappa ')^{X}$
into $ w (1,L^{2}\kappa ')$, which brings the decay rate $\kappa $
back to better than the original value. In particular we can choose
$\bar{w} = w ^{2}$ for $L>2$. To summarize: for $\lambda $ small, the
rescaling more than restores the loss of decay caused by convolution
so that the whole RG map will actually strengthen this parameter in the
norm.

\begin{lemma} \label{lem:integrability1} (Integrability of $e^{-\vhat}$):
If $M$ is a polynomial form with degree $m$, then 
\[
\|M e^{-\vhat} \|_{w,h}
\le c_{m} (h/\fh )^{m}|M|_{\fh},
\hspace{2mm} |f e^{\vhat }|_{\fh } \le c |f|_{\fh }.
\]
\end{lemma}

\Proof{ of second estimate } $|f e^{\vhat }|_{\fh } \le |f | _{\fh }
e^{|\vhat |_{\fh }} $ and $|\vhat |_{\fh } \le c|\vhat |_{h} \le c$ by
hypotheses. \qed

The first estimate is reduced to the case $h=1$, $\lambda = c$ by
scaling $\phi = h\phi '$. See the proof of (7.4) in
\cite{BEI92}. Similarly we can bound the $\exp (-\nu \tau )$ in $\exp
(-\vhat )$ for $\nu $ either positive or negative, but we must then
have a weight that allows growth at infinity as in

\begin{lemma} \label{lem:integrability2} (Integrability of $e^{\nu \tau }$):
$\|e^{-\nu \tau }\|_{w^{-1},h} \le c$ and $|\exp (-\nu \tau )|_{\fh }
\le c$.
\end{lemma}

\begin{lemma} \label{lem:interaction-scales-down} (Interaction scales
down): If $r$ is normalized, then $\|r \exp (- \vhat) \|_{w,h} \le c L
^{-6} \|r\|_{w,\cS h/2}$ and $|r|_{\fh '} \le c [\fh '/(\fh - \fh
')]^{6}|r|_{\fh }$ for $\fh ' < \fh $. More generally, if
(c.f. Definition~\ref{def:normalization}), $r_{t} = O (t^{p})$, then
$L^{-6}$ is replaced by $L^{-p}$.
\end{lemma}

\proof{} By Taylor's theorem in $t$ applied to $r_{t} = r (t\phi )$ it
suffices to estimate the $w,h$ norm of
\[
	\int _{0}^{1} \, \frac{(1-t)^{5}}{5!}
	\frac{d^{6}}{dt^{6}} r_{t}e^{- \vhat} \,dt ,
\]
which is bounded by $c$ times terms of the form $ \sup_{t} \|
r^{(6)}_{t} M e^{- \vhat}\|_{w,h}$ where $r^{(6)}$ denotes a sixth $\phi
,\psi $ derivative of $r$ and $M$ is a monomial in $\phi ,d\phi $ of
degree $6$.  This is less than
\begin{align*}
	\sup_{t}
	\| r^{(6)}_{t}\|_{w_{t},h}\, 
	\|M e^{- \vhat}\|_{w/w_{t},h}
&=
	\sup_{t}
	\| r^{(6)}\|_{w,th}\, 
	\|M e^{- \vhat}\|_{w/w_{t},h}\\
&
\le
	\| r^{(6)}\|_{w,h}\, 
	\|M e^{- \vhat}\|_{w,h} ,
\end{align*}
where $w_{t}= w (t\phi )$, because the norm is decreasing in $w$ and
$w/w_{t} \ge w$.  $r^{(6)}$ is bounded by $c(\cS h)^{-6}$ times the
$w,\cS h/2$ norm by Lemma~\ref{lem:properties-of-norm} (iv) and
$Me^{-\vhat}$ is bounded by $h^{6}$ using
Lemma~\ref{lem:integrability1}.  The second inequality is proved by
the same steps with the $\fh $ norms.  \qed

\begin{lemma} \label{lem:extraction} (Extraction): The solution $(\nu ',
v',r')$ to
\begin{equation}\label{eq:defines-prime}
e^{-\vhat} + \rhat = e^{ - \nu '\tau } (
e^{-v'} + r '), \hspace{2mm} (v',r') \text{ normalized}
\end{equation}
satisfies
\renewcommand{\theenumi}{(\roman{enumi})} 
\begin{enumerate}
\item[(i)] $|\vhat - \nu '\tau - v'|_{\fh } \le c |\rhat|_{\fh}$
\item [(ii)]$|\nu ' - \nu |\fh ^{2} + |\lambda '- \lambda | \fh ^{4} \le c
|\rhat|_{\fh} $ where $\gamma =0$ in $\rhat $
\item[(iii)] $\|r '\|_{w,h} \le c \|\rhat\|_{w^{2},h}$
\item[(iv)] $|r '|_{\fh } \le c|\rhat|_{\fh}$
\end{enumerate}
provided that for each estimate the right-hand side is less than a
constant $c$ which is small.
\end{lemma}

\Proof{ of (i)} Define $|\cdot|_{\fh }^{(p)}$ by truncating the sum in
the definition of $|\cdot |_{\fh }$ so that only derivatives of order
$p$ or less are included. This seminorm satisfies $|F_{1}F_{2}|_{\fh
}^{(p)} \le |F_{1}|_{\fh }^{(p)} |F_{2}|_{\fh }^{(p)}$. Rewrite the
$(\nu ', v',r')$ equation,
\begin{equation}\label{eq:proof-of-part-i}
\vhat  - \nu '\tau - v '= 
\ln \left(
1 + e^{\vhat}\rhat - e^{\vhat-\nu '\tau }r '
\right).
\end{equation}
Expand the $\ln $, take the $|\cdot |_{\fh }^{(p)}$ with $p=4$:
\[
|\vhat  - \nu '\tau - v '|_{\fh }^{(p)} \le \sum _{j} \frac{x^{j}}{j}, 
\hspace{2mm}
x = |e^{\vhat}\rhat|_{\fh }^{(p)} + 
|e^{\vhat-\nu '\tau }|_{\fh }^{(p)} |r'|_{\fh }^{(p)}. 
\]
It suffices to prove the estimate for $p = 4$ because $|\vhat - \nu
'\tau - v '|_{\fh }^{(4)} = |\vhat - \nu '\tau - v '|_{\fh }$ whereas
$|\rhat |_{\fh }^{(p)} \le |\rhat |_{\fh }$. $x =
|e^{\vhat}\rhat|_{\fh }^{(p)}$ because the first five derivatives of
$r'$ are zero by normalization.  Also, $|e^{\vhat}\rhat|_{\fh }^{(p)}
\le c |\rhat |_{\fh }$ as in the proof of
Lemma~\ref{lem:integrability1}. By hypothesis, the sum is dominated by
the first term so that
\[
|\vhat  - \nu '\tau - v '|_{\fh }^{(p)} \le
c |\rhat|_{\fh }.
\]

\Proof{ of (ii)} Bound the left-hand side by $ c |\vhat -
\nu ' \tau - v'|_{\fh }$ and use part (i).

\Proof{ of (iii)} Let 
\begin{equation}\label{eq:rprime-alpha}
r '(\alpha ) = e^{\nu '\tau } (e^{-\vhat } + \alpha \rhat ) - e^{-v'
(\alpha )}
\end{equation}
be the solution when $\rhat $ is replaced by $\alpha \rhat $.  For
$\alpha $ in the disk $\|\alpha \rhat \|_{w^{2},h} \le c_{1}$, with
$c_{1}$ small, { we show that} $\|r' (\alpha ) \|_{w,h}$ is bounded by $c${.  First, }
$\exp (\nu '\tau - \vhat )$ and $\exp (-v' )$ satisfy a $(\vhat ,w)$
hypothesis by (a) parts (ii) and (i) with $\fh = h${, and }  (b) $|\rhat |_{h} \le
\|\rhat \|_{w^{2},h}${. Second,}
\[
\|\exp (\nu '\tau ) (\alpha \rhat ) \|_{w,h} \le
\|\exp (\nu '\tau ) \|_{w^{-1},h} \|\alpha \rhat  \|_{w^2,h}
\le c ,
\]
using Lemma~\ref{lem:integrability2}. { Since} $r' (\alpha )$
vanishes at $\alpha =0$ { we have}, by analyticity in $\alpha $,
\[
\|r' \| _{w,h} \le c\|\rhat \|_{w^{2},h}.
\]

\Proof{ of (iv)} By taking the norm of (\ref{eq:rprime-alpha}),
\[
|r '(\alpha ) |_{\fh } \le e^{|\nu ' \tau |_{\fh }} ( e^{|\vhat |_{\fh
}} + |\alpha \rhat |_{\fh }) + e^{|v'|_{\fh }} \le c.
\] 
Therefore, as in part (iii), $|r'|_{\fh } \le c |\rhat |_{\fh } $.\qed

Define 
\begin{gather}\label{eq:rg0}
\rmain = \text{constant and linear in $\gamma $ part of } 
\cS \mu _{C} \ast ( e^{-v})^{\cG _{1}} - e^{ - \vhat} ,
\end{gather}
with $\vhat = \vt + \nut \tau $ as in
Proposition~\ref{prop:recursion1}. The right-hand side is expanded in
a series in powers of $\gamma $ and truncated at order $\gamma
$.  For computing the observable, we only need the derivative in
$\gamma $ at zero so there is no loss in such truncation.

\begin{lemma}\label{lem:r-main-estimates} (Remainder after
perturbation theory is small in norm): Suppose that $\sum _{y}
C_{xy}=0$ and $|\fh | \le c_{L}\|C \|^{\frac{1}{2}}$.  Then
\[
	\|\rmain \|_{w,h} \le c_{L} \|C \|^{2}|\lambda |,
\hspace{3mm}
	|\rmain |_{\fh}
\le
	c_{L} \|C \|^{3} |\lambda |^{3}.
\]
\end{lemma}

\newcommand{\barJ}{\bar{J}}
\newcommand{\underbarJ}{\underline{J}}

The main obstacle to proving this lemma is the ``large field problem''
described at the end of
Section~\ref{section:second-order-perturbation-theoryI}.  Using the
notation of Lemma~\ref{lem:second-order-perturbation}, let $\Vhat
_{t}= -V_{t}+Q_{t}$ with $Q$ linearized in $\gamma $ and let $E_{q}
(A) = \sum _{j=0}^{q}A^{q}/q!$ denote the exponential truncated at
order $q$.  We solve the large field problem by splitting $\Vhat _{t}
= \barJ_{t} + \underbarJ_{t}$ and using the approximation $\exp
(\underbarJ_{t})E_{q} (\barJ_{t})$ in place of $\exp (\Vhat _{t})$.
$\barJ_{t}$ will contain the polynomial of degree six which would ruin
integrability if fully exponentiated.

$Q_{t}$ is a sum of terms of the form $P_{x} C^{j} _{x,y}P_{y} + O
(\gamma )$, where $P_{x}$ is a polynomial in $\phi _{x}, \psi _{x}$
and conjugates.  Let $\underbar{Q}_{t}$ be the same sum with $P_{y}$
replaced by $P_{x}$.  The degree 6 term that seems to be in
$\underbar{Q}_{t}$ is actually zero because $\sum _{y}C_{xy} =0$.  We
choose $\underbarJ _{t} = -V_{t} + \underbar{Q}_{t}$.  Then $\barJ_{t}
= Q_{t} - \underbar{Q}_{t}$ has the property that $\cS \barJ_{t}=0$
and
\begin{equation}\label{eq:vtilde-from-J}
\vhat = \cS \Vhat _{t=1} = -\cS \underbarJ _{t=1}.
\end{equation}
Since $\lambda $ is small we can assume the same $(\vhat ,w)$ hypothesis
for $v$ and $\vhat $. 

Note that $\underbarJ_{t}$ contains the $\tau ^{2}$ terms. We keep
them in a standard exponent because $\exp (-\lambda \tau ^{2} )$ is
useful for suppressing large values of $\phi $, unlike $E_{q}
(-\lambda \tau ^{2})$.

\Proof{ of Lemma~\ref{lem:r-main-estimates}} Equation (\ref{eq:vtilde-from-J})
implies that at $t=1$, $\cS \exp (\underbarJ_{t})E_{q} (\barJ_{t}) =
e^{-\vhat}$. Recall from
Section~\ref{section:second-order-perturbation-theoryI} the Duhamel
formula
\begin{equation}\label{eq:rmain1}
	\rmain
=
	\cS \int _{0}^{1} \,dt \, \mu _{(1-t)} 
	\ast \cL e^{\underbarJ_{t}}E_{q} (\barJ_{t}).
\end{equation}
By a calculation based on the differentiation rule $D E_{q} (\barJ ) =
E_{q-1} (\barJ)D\barJ$ ,
\begin{gather}\label{}
	\cL e^{\underbarJ}E_{q} (\barJ)
=
	e^{\underbarJ}E_{q-1} (\barJ)\left[
	\cL \Vhat  - \frac{1}{2}\Vhat \join \Vhat 
	\right] + \nonumber\\
	\frac{1}{q!} e^{\underbarJ} \barJ^{q}\left[
	\cL \underbarJ - 
	\frac{1}{2}\underbarJ\join \underbarJ
	\right] +
	\frac{1}{2(q-1)!} e^{\underbarJ} \barJ^{q-1}
	\barJ\join \barJ.
\end{gather}
We have omitted $t$ subscripts.  We choose $q=2$.  As in
(\ref{eq:pert1}),
\begin{align}\label{eq:rmain2}
	\cL e^{\underbarJ}E_{2} (\barJ)
&=
	e^{\underbarJ}E_{1} (\barJ)\left[
	Q\join V - \frac{1}{2}Q\join Q
	\right] + \nonumber\\
&
	\frac{1}{2}e^{\underbarJ}\barJ^{2}\left[
	\cL \underbarJ - \frac{1}{2}\underbarJ\join \underbarJ
	\right] +
	\frac{1}{2}e^{\underbarJ}\barJ
	\barJ\join \barJ.
\end{align}
Consider the term $\exp(\underbarJ )E_{1} (\barJ)Q_{t}\join V_{t}$ in
(\ref{eq:rmain2}). Since $E_{1} (\barJ) = 1 + \barJ $ it splits into
two terms, one of which is $F:=\exp(\underbarJ )Q_{t}\join V_{t}$.  $F$
is a sum of products of terms of the form $( Me^{-v})^{\cG_{1} }$ as
in Lemma~\ref{lem:integrability1} because there are three sites in
$\cG _{1}$ where there are monomials corresponding to the three
factors $v_{x}, v_{y},v_{z}$ in $F$. If we consider the $\lambda \tau
^{2}$ parts of $v$ then we find that the total degree of the monomials
is $3 \times 4 - 4 = 8$.  The $-4$ is because $F$ contains two $\join
$, one explicit and one in $Q_{t}$.  Associated with these two $\join
$ are $C_{x,y}$ and $C_{y,z}$. There is also $\lambda ^{3}$.
Therefore, by Lemmas \ref{lem:properties-of-norm} -
\ref{lem:integrability1}, $\|( Me^{-v})^{\cG_{1} }\|_{w,h}$ is bounded
by $c_{L} h^{8}|\lambda |^{3}\|C \|^{2} = c_{L} |\lambda |\|C
\|^{2}$. The $c_{L}$ includes factors $L^{4}=|\cG_1|$ from sums over
$x,y,z$.

The other term in $\exp(\underbarJ )E_{1} (\barJ)Q_{t}\join V_{t}$ is
$\barJ F$. This is smaller by an additional factor $c_{L}h^{6}|\lambda
|^{2}\|C \|^{2}$ by the same methods.  In fact, all of the terms in
(\ref{eq:rmain1},\ref{eq:rmain2}) can be estimated by this procedure
by $c_{L} |\lambda |\|C \|^{2}$. The $O (\gamma )$ terms may be
absorbed into these bounds because, according to the $(\vhat ,w)$
hypothesis, $|\gamma b| \le c|\lambda |$.

The $|\cdot |_{\fh}$ norm is estimated in the same way but taking into
account the hypothesis on $\fh $, (i) $|\lambda |^{3}\fh ^{p} \le
c_{L,p}|\lambda |^{3}$ and (ii) if $M$ is a monomial of order $2p$
then $|M|_{\fh } \le c\fh ^{2p} \le c_{L,p}\|C \|^{p}$. There are at
least two factors of $C$ in every term and for terms with only two at
least one more is contributed by (ii). \qed

\emph{Hypothesis ($\lambda ,r$ small depending on $L$):} $|\lambda |, \|r
\|_{w,h} \text{ and } ( h/\fh )^{4}|r |_{\fh} \le c_L$ where $c_{L}$
is determined in the proof of the next theorem.

\begin{proposition} \label{prop:control-of-r} Let $v$ be given by
(\ref{eq:defines-v}), $h, h' = |\lambda |^{-1/4}, |\lambda
'|^{-1/4}$ and $c_{L}\sqrt{\|C \|}\ge \fh \ge \fh ' \ge L\sqrt{\|C
\|}$.  Then the map from $(v,r)$ to $(v',r',\nu ')$ defined by
\[
	\cS \mu \ast (e^{-v}+r)^{\cG_{1} }
=
	e^{ - \nu ' \tau } (
	e^{-v'} + r'
	) ,
\]
with $v', r'$ normalized, satisfies
\begin{align*}
&(i)\hspace{2mm}
\|r' \|_{w,h'} \le 
c_{L} |\lambda | \|C \|^{2} + c L^{-2} \|r \|_{w,h} 
\\[2mm]
&(ii)\hspace{2mm}
|r' |_{\fh '} \le 
c_{L} |\lambda |^{3} \|C \|^{3} + 
c L^{-2} (\fh '/\fh )^{6}|r |_{\fh} + c_{L}\|C \|^{5}h^{-10}\|r\|_{w,h }
\\[2mm]
&(iii)\hspace{2mm}
|\beta ' - \betat | \fh ^{2} + |\lambda '-\lambdat | \fh ^{4} \le 
c_{L} |\lambda |^{3} \|C \|^{3} + c L^{-2} (\fh '/\fh)^{6} |r |_{\fh} \\[2mm]
&\hspace{50mm} + c_{L}h^{-10}\|C \|^{5}\|r\|_{w,h }.
\end{align*}
\end{proposition}

\medskip

\noindent \Proof{ of (i) and (ii)} By solving $\T (\exp
(-v)+r)^{\cG_1} = \exp (-\vhat ) + \rhat $ for $\rhat $,
\begin{gather}
\rhat = \rmain 
+ \cS E_{q-1} (\Delta ) \sum _{x \in \cG _{1}}
(e^{-v})^{\cG _{1}\setminus \{x\}}r_{x} \nonumber\\
+ \cS E_{q-1} (\Delta ) \sum _{X\subset \cG _{1}, |X|>1} 
(e^{-v})^{\cG _{1}\setminus X}r^{X} \nonumber\\
+ (\T -\cS E_{q-1} (\Delta )) 
\sum _{X\subset \cG _{1}} 
(e^{-v})^{\cG _{1}\setminus X}r^{X}. \label{eq:rhat-formula}
\end{gather}

\medskip \noindent 
\emph{First term in (\ref{eq:rhat-formula}):} By
Lemma~\ref{lem:r-main-estimates} $ \|\rmain \|_{w^{2},h}\le c_{L}\|C
\|^{2 } |\lambda |$.

\medskip \noindent \emph{Second term in  (\ref{eq:rhat-formula}):} We
choose $q=1$. Let $X = \cG _{1}\setminus \{x \}$. $\cS (\exp (-v
))^{X}$ equals $\exp (-v_{\text{new}} )$ where $v_{\text{new}}$ still
satisfies the $(\vhat , w)$ hypothesis\footnote{because it equals
$\lambda _{\text{new}}\tau ^{2} + \text{ observable}$ with $\lambda
_{\text{new}} = \lambda (1-L^{-4})$.}.  By
Lemma~\ref{lem:interaction-scales-down},
Lemma~\ref{lem:properties-of-norm} part (iii) and $w^{2} \ge \cS w
^{\cG _{1}}$,
\begin{align}\label{eq:one-r-term}
\|\sum _{x \in \cG _{1}}
\cS (e^{-v})^{X} \cS r_{x} 
\|_{w^{2}, 2h} 
&\le cL^{4} L^{-6} \|\cS r\|_{w^{2},\cS h}
\le cL^{-2} \|r\|_{w,h}.
\end{align}

\medskip \noindent
\emph{Third term in  (\ref{eq:rhat-formula}):} By
Lemmas~\ref{lem:properties-of-norm} and \ref{lem:integrability1}, $\|
\text{third term}\|_{w^{2},2h} \le c_{L}\|r\|^{2}_{w,h}$.

\medskip \noindent \emph{Fourth term in  (\ref{eq:rhat-formula}):} By
Lemma~\ref{lem:fluct-equals-scaling}, with $\bar{w} = w^{2}$ as
explained below the $(\vhat ,w)$ hypothesis, followed by
Lemmas~\ref{lem:properties-of-norm} and \ref{lem:integrability1},  $\|
\text{fourth term}\|_{w^{2},2h} \le
h^{-2} \|r\|_{w,h}$ because there are less than $c_{L}$ terms in the
sum over $X$ and each has one or more factors of $r$ and $q=1$.

Therefore
\begin{equation}\label{eq:rhat-bound1}
\|\rhat \|_{w^{2},2h} \le 
c_{L} |\lambda | \|C \|^{2} + c L^{-2} \|r \|_{w,h} ,
\end{equation}
where we used the hypotheses ($r$ and $h^{-2} = \sqrt{|\lambda |}$
small depending on $L$) to put the estimates for the third and fourth
terms inside $cL^{-2} \|r\|_{w,h}$. 

To estimate the $\fh '$ norm, we use (\ref{eq:rhat-formula}) with $q=5$.
Consider the second term in (\ref{eq:rhat-formula}). Let 
\[
F = \sum _{x \in \cG _{1}} (e^{-v})^{\cG _{1}\setminus \{x\}}r_{x}.
\]
Note that $F_{t} = O (t^{6})$ is normalized, using the subscript $t$
notation of Definition~\ref{def:normalization}. However, $(\Delta
^{n}F)_{t} = O (t^{6-2n})$. By
Lemma~\ref{lem:interaction-scales-down}, with $\fh $ replaced by $L\fh
/2$, and Lemma~\ref{lem:properties-of-norm} part (iii)
\[
|\cS \Delta ^{n}F |_{\fh '} \le 
c \left[\fh '/ ( L\fh )\right]^{6-2n}|\Delta ^{n}F|_{\fh /2} ,
\]
for $n=0,1,2$. When $n=3,4,5$ we still have this bound (by
Lemma~\ref{lem:properties-of-norm} (iii)) because $\fh '/ ( L\fh )
\le 1$.  By Lemma~\ref{lem:integrability1}, $|\exp (-v)|_{\fh } \le 1
+ L^{-4}$ for $|\lambda | < c_{L}$.  Therefore, by
Lemma~\ref{lem:properties-of-norm},
\[
|\cS \Delta ^{n}F |_{\fh '} \le 
c L^{4} \left[\fh '/ ( L\fh )\right]^{6-2n} \left[\|C \|\fh^{-2}
\right]^{n} |r|_{\fh } ,
\]
and then, by  $\fh ' \ge L\sqrt{\|C \|}$,
\[
|\text{second term in (\ref{eq:rhat-formula})} |_{\fh '} \le c L^{-2}
[\fh '/ \fh ]^{6} |r|_{\fh }.
\]
The third term in (\ref{eq:rhat-formula}) is absorbed into the same
bound by using the same argument. There are merely more factors of $r$
which make it much smaller by the hypothesis on $|r|_{\fh }$. The
fourth term in (\ref{eq:rhat-formula}) is bounded by
Lemma~\ref{lem:fluct-equals-scaling} so that
\begin{equation}\label{eq:rhat-bound2}
|\rhat |_{\fh '} \le c_{L} |\lambda |^{3} \|C \|^{3} + c L^{-2} (\fh
'/\fh )^{6}|r |_{\fh} + c_{L}\|C \|^{5}h^{-10}\|r\|_{w,h }.
\end{equation}
The proof of (i) and (ii) is completed by applying
Lemma~\ref{lem:extraction} to (\ref{eq:rhat-bound1},
\ref{eq:rhat-bound2}).

\medskip \noindent 
\Proof{ of (iii)}  Immediate consequence of Lemma~\ref{lem:extraction} and
(\ref{eq:rhat-bound1},\ref{eq:rhat-bound2}).
\qed 

\medskip

\Proof{ of Proposition~\ref{prop:beta-lambda-recursion}} Recall that
the \emph{finite} volume Green's function $G_{\lambda _{0}}^{\Lambda
_{0}} (\beta _{0},x) $ has been expressed as an integral,
(\ref{eq:green-and-bzero}). This integral is convergent for all $\beta
_{0} \in \CC$, provided the real part of $\lambda _{0}$ is positive,
which is the case for $\lambda _{0} \in \Dbarlambda $.  In this case
we can also define $(\beta _{j},\lambda _{j},r_{j})$ by conglomerating
the $j$ RG steps into one giant step (for RG), with covariance
equal to the sum of all the covariances of the individual RG steps,
and applying it to the initial interaction.  The big RG step is a
convolution of a Gaussian into an interaction that decays faster than
any Gaussian so it is convergent and the result defines $(\beta
_{j},\lambda _{j},r_{j})$. Furthermore, $r_{j}$ is entire in $\phi $.
 The sequence $(\beta _{j},\lambda _{j},r_{j})$ is defined for all $j$
by the remark below (\ref{eq:coordinates2}).

In order to obtain useful estimates on the sequence, we rotate the
contours of integration of each $\phi _{x}$ (both real and imaginary
parts). For any $|\theta | \le b_{\beta } + b_{\lambda }/4 + 9\epsilon
/8 - \pi /2$ we may rotate $\phi _{x} \rightarrow \exp (i\theta /2)
\phi _{x}$ and $\phibar _{x} \rightarrow \exp (i\theta /2)\phibar
_{x}$. Likewise $\psi _{x} = (2\pi i)^{-1/2}d\phi _{x}$ and $\psibar
_{x} = (2\pi i)^{-1/2}d\phibar _{x}$ pick up factors $\exp (i\theta
/2)$.  The effect of the rotation is to replace $\beta ,\lambda ,A$
with
\[
\beta (\theta ) = e^{i\theta }\beta , \hspace{2mm} 
\lambda (\theta ) =  e^{2i\theta }\lambda   , \hspace{2mm} 
A (\theta ) = e^{i\theta }A.
\]
Since the covariance $A^{-1}_{xy} = U^{\Lambda } (\beta ,x-y)$ is
determined by $\Gamma $ through (\ref{eq:dirichlet-scale-decomp}), the
rotation of $A$ is equivalent to an anti-rotation $\Gamma (\beta )
\rightarrow \Gamma (\theta ) = \Gamma (\beta )\exp (-i\theta
)$. Recalling the discussion of $b_{\beta },b_{\lambda }$ below
(I.1.15), the above condition on $\theta $ is sufficient to ensure 
that $|\arg \lambda (\theta ) | < \pi /2 - \epsilon /4$ { for 
any $|\arg\lambda|<b_{\lambda}+\epsilon$, since} by
assumption,
\[
2 (b_{\beta } + \epsilon ) + \frac{3}{2} (b_{\lambda } + \epsilon ) <
\frac{3}{2}\pi.
\]
Hence the integrals remain rapidly convergent at $\infty $ and there
is no contribution from the contours at $\infty $. We obtain the
functional equation
\begin{equation}\label{eq:functional-equation-for G}
G_{\lambda ,\Gamma }^{\Lambda} (\beta ,x) = G_{\lambda (\theta
),\Gamma (\theta ) }^{\Lambda } (\beta (\theta ),x) e^{i\theta }.
\end{equation}

The sequence $(\beta _{j},\lambda _{j},r_{j})$ transforms to $(\beta
_{j} (\theta ),\lambda _{j} (\theta ),r_{j} (\theta ))$, where $\beta
_{j} (\theta ) = \beta _{j}\exp (i\theta )$, $\lambda _{j} (\theta ) =
\lambda _{j}\exp (2i\theta )$, and $r_{j} (\theta ,\phi ) = r_{j}
(\phi \exp (i\theta ))$. It is evident that $\beta _{j}, \lambda _{j}$ 
transform in the same manner as $\beta ,\lambda $ because shifts in
the coupling were defined in Section~\ref{section:coordinates} as
derivatives of functions of $\phi $, now $\phi \exp (i\theta /2)$ and
each derivative introduces a factor $\exp (i\theta /2)$.

Define a decomposition $\Dbarb (\frac{1}{2}) = H^{+}\cup H^{-}$ of the
$\beta $ domain into two overlapping fattened sectors. Here
\[
H^{\pm} = \{\beta \not =0: |\arg\beta \mp \theta | < \frac{\pi
}{2} - \frac{\epsilon }{8} \} + \cB (\frac{1}{2}),
\]
where $\theta  = b_{\beta } + b_{\lambda }/4 + 9\epsilon /8 - \pi /2$.

\medskip \emph{Claim 1: } Suppose that $\beta _{0}$ is in $H_{-}$ and
$\lambda _{0}$ is in $\Dbarlambda $.  Let $\fh _{j} = L \|\Gamma
(\beta _{j-1}) \|^{1/2}$. For $L$ sufficiently large, there is a
constant $k_{L}$ such that
\begin{align}
&\|r_{j} (\theta ) \|_{w,h_{j}} 
\le k_{L} |\lambda _{j-1}| \|\Gamma (\beta _{j-1}) \|^{\frac{1}{2}}\nonumber\\
&|r_{j} (\theta ) |_{\fh _{j}} 
\le k_{L}  |\lambda _{j-1}| ^{3} \|\Gamma (\beta _{j-1}) \|^{3},
\label{eq:r-theta-claim}
\end{align}
for all $j\ge 1$ such that $(\beta _{j-1}, \lambda _{j-1})$ lies in
$H_{-} \times \Dbarlambda $.  The analogous claim with $H_{-}$
replaced by $H_{+}$ and $r _{j}(\theta )$ replaced by $r _{j}(-\theta
)$ is also true.

\medskip

\Proof{ by induction} Since $\lambda _{j-1} \in \Dbarlambda $, we have
as above that $|\arg\lambda _{j-1} (\theta )| < \pi /2 - \epsilon /8$
and so the $\lambda $ hypothesis is valid for $\lambda _{j} (\theta
)$. The covariance is
\[
\Gamma (\theta ) = \Gamma (\beta _{j-1})e^{-i\theta } = 
\frac{e^{-i\theta }}{1+\beta _{j-1}}\Gamma (0) = 
\frac{e^{-i (\theta + \arg (1+\beta _{j-1}))}}{|1+\beta
_{j-1}|}
\Gamma (0),
\] 
and we verify the $\mu $ hypothesis. First we observe that $\Gamma
(0)$ is positive semi-definite.  Second, since $b_{\beta }+ \epsilon
< 3\pi /4$, we have that $|1+\beta _{j-1}| > (\sqrt{2}-1)/2$. Third,
our definition of $H^{-}$ guarantees that $|\theta + \arg
(1+\beta _{j-1})| < \pi /2 - \epsilon /8$.

As a result, Proposition~\ref{prop:control-of-r} is applicable to the
rotated parameters.  By part (ii),
\[
|r _{j+1}(\theta ) |_{\fh_{j+1}} \le 
\frac{1}{2} k_{L} |\lambda _{j}|^{3} \|\Gamma (\beta _{j} ) \|^{3}
+ \frac{1}{4} (\fh _{j+1}/\fh _{j})^{6} |r_{j} (\theta )
|_{\fh _{j}} 
+ c_{L}\|\Gamma (\beta _{j})\|^{5}h^{-10} \|r_{j} (\theta )\|_{w,h_{j}
}.
\]
We made a $j$ independent choice of $L$ large to get the second term
in this particular form. The first term has the constant $k_{L}$
because we choose $k_{L}$ in the inductive hypothesis to make it
so. This bound proves the second inequality of claim 1 for $j=1$
because $r_{0}=0$. In the right-hand side, substitute the inductive
assumptions (\ref{eq:r-theta-claim}) to obtain
\[
|r_{j+1} (\theta ) |_{\fh _{j+1}} 
\le k_{L}\left(
\frac{1}{2}  |\lambda _{j}|^{3} \|\Gamma (\beta _{j}) \|^{3}
+ \frac{3}{8}  |\lambda _{j-1}|^{3}
\|\Gamma (\beta _{j}) \|^{3}\right) ,
\]
where $c_{L}\|\Gamma (\beta _{j})\|^{5}h^{-10} \|r_{j} (\theta
)\|_{w,h_{j} }$ went into the second term by a $j$ independent choice
of $\Dbarlambda $ so that $|\lambda |$ is small, making $h$ large and
consequently $h _{j}^{-10} \ll |\lambda _{j}|^{2}$.  
By the $\lambda $ recursion $\frac{3}{8}|\lambda _{j-1}| \le
\frac{4}{8}|\lambda _{j}|$. Thus we have advanced $j$ to $j+1$ in the
second estimate estimate of claim 1.
The advance of $j$ to $j+1$ in the first estimate is proved
similarly. But first observe that there exists an $L$-independent
constant $k$ such that 
\[
\frac{\|\Gamma (\beta _{j-1}) \|}{\|\Gamma (\beta _{j}) \|} =
\left|
\frac{1+\beta _{j}}{1+\beta _{j-1}}
\right| =
\left|
\frac{1+\beta _{j}}{1+L^{-2}\beta _{j}+O (\lambda _{j-1})}
\right| \le
k^{2}L^{2}
\]
for all $\beta _{j} \in \Dbarb (1/2)$. (The relation between $\beta
_{j-1}$ and $\beta _{j}$ follows from
Proposition~\ref{prop:control-of-r} (iii) and the inductive assumption
(\ref{eq:r-theta-claim})). Hence, by
Proposition~\ref{prop:control-of-r} (i) and
(\ref{eq:r-theta-claim}), we have
\[
\|r_{j+1} (\theta ) \|_{w,h_{j+1}} \le
{ k}_{L} \|\Gamma (\beta _{j}) \|^{\frac{1}{2}} 
\left(
\frac{1}{2}|\lambda _{j}| + cL^{-2}kL|\lambda _{j-1}| 
\right) ,
\] 
so we may choose $L>{ 4}ck$ to complete the induction.

For later use we observe that if, in the above argument, we use
(\ref{eq:rhat-bound1}, \ref{eq:rhat-bound2}) in place of
Proposition~\ref{prop:control-of-r} (i) and (ii), we obtain

\medskip \emph{Claim $1'$:} (\ref{eq:r-theta-claim}) holds when $r_{j+1}$
is replaced by $\rhat _{j}$.

\medskip
\emph{Claim 2:} Let $(\beta _{j},\lambda _{j})$ with $j=0,1,\dots
,M-1$ be an RG sequence in $\Dbarb (1/2) \times \Dbarlambda $. If
$\beta _{j}$ lies in $H_{-} \setminus H_{+}$, then $\beta _{j+1}$ is
not in $H_{+}$.

\medskip
The idea is that the factor $L^{2}$ in the $\beta $ recursion causes a
large expansion in $\beta $ in a direction that takes $\beta _{j}$
further from $H_{+}$.  There are other terms in the $\beta $
recursion, but they are small by claim 1.

\proof{} By the hypothesis, $\Re \beta _{j} (-\theta -\epsilon /8) \le
-\frac{1}{4}$. Therefore, $\Re L^{2}\beta _{j} (-\theta -\epsilon /8)
\le -\frac{1}{4} L^{2}$, so no $\beta $ within distance one
of $L^{2}\beta _{j} $ can be in $H_{+}$. On the other hand, by claim 1
and part (iii) of Proposition~\ref{prop:control-of-r}, $|\beta _{j+1}
(\theta ) - \betat _{j+1}(\theta )| \ll 1$. Therefore $|\beta _{j+1} -
L^{2}\beta _{j}| \ll 1 $. Consequently $\beta _{j+1} $ is not in
$H_{+}$.

By claim 2, either the whole sequence $(\beta _{j},\lambda _{j})$ with
$j=0,1,\dots ,M-1$ lies in $H_{-}$ or it lies in $H_{+}$ or in
both. Therefore Proposition~\ref{prop:beta-lambda-recursion} follows
from claim 1 applied with either $H_{+}$ or $H_{-}$,
Proposition~\ref{prop:control-of-r} (iii), along with
(\ref{eq:recursion-beta-lambda}). \qed

\emph{Remark:} The functional equation (\ref{eq:functional-equation-for
G}) provides an analytic continuation of $G_{\lambda }$ into the
larger domain $|2 \arg\beta  -(3/2) \arg \lambda | < 3\pi
/2$, $|\arg \beta |<\pi $, $|\arg \lambda | < \pi $, even though the
defining integral fails to converge for $|\arg \lambda |> \pi
/2$. However, to obtain the end-to-end distance from $G_{\lambda }$,
we require $b_{\beta }> \pi /2$ and so $b_{\lambda }< \pi /3$.

Proposition~\ref{prop:error-for-b} involves derivatives with respect
to the observable parameter $\gamma $ at $\gamma =0$.  Derivatives at
zero will be denoted by $\gamma $ subscripts as in $v_{\gamma }$ and
$r_{\gamma }$. We will repeatedly use four principles: (1)
distinguishing \emph{partial} dependences on $\gamma $ by subscripts
on $\gamma $;  (2) the total $\gamma $ derivative is the sum of the
partial derivatives with respect to partial dependences; (3) Functions
of $\gamma$  may be replaced by linearizations{; and} (4) Cauchy
estimates on derivatives. These are illustrated in detail in the proof
of Corollary~\ref{cor:extraction} and are subsequently used without
comment.

\begin{corollary} \label{cor:extraction} (to
Lemma~\ref{lem:extraction}) 
\renewcommand{\theenumi}{(\roman{enumi})}
\begin{enumerate}
\item[(i)] 
$|\vhat _{\gamma } - v'_{\gamma }|_{\fh } \le c \left(|\rhat |_{\fh }
|\vhat _{\gamma }|_{\fh } + |\rhat _{\gamma }|_{\fh} \right)$ 
\item[(ii)] $|r'_{\gamma }|_{\fh } \le c \left(|\rhat _{\gamma }|_{\fh } +
|\rhat |_{\fh } |\vhat _{\gamma }|_{\fh } \right)$
\item[(iii)] $\|r'_{\gamma } \|_{w,h} \le c \left( \|\rhat _{\gamma }
\|_{w^{2},h} + \|\rhat  \|_{w^{2},h}|\vhat _{\gamma }|_{h} \right)$
\end{enumerate}
\end{corollary}

\Proof{ of (i) } Firstly, $\vhat - \nu '\tau - v' $ is a function of
$\gamma _{\vhat }$ when $\vhat $ is replaced by $\vhat + \gamma
_{\vhat }\vhat _{\gamma }$. It is $\fh $ norm analytic and bounded by
$c$ uniformly in the disk $|\gamma _{\vhat }\vhat _{\gamma }|_{\fh } \le c$. By
the Cauchy formula for the derivative at $\gamma =0$ and
Lemma~\ref{lem:extraction} part (i), $|(\vhat - \nu '\tau -
v')_{\gamma _{\vhat }}|_{\fh } \le c|\rhat |_{\fh } |\vhat _{\gamma
}|_{\fh } $.  Secondly, $\vhat - \nu '\tau - v' $ is a function of
$\gamma _{\rhat }$ when $\rhat $ is replaced by $\rhat + \gamma
_{\rhat }\rhat _{\gamma } $. It is norm analytic and, by
Lemma~\ref{lem:extraction} part (i), is bounded by $c$ uniformly in
the disk $|\gamma _{\rhat }\rhat _{\gamma }|_{\fh } \le |\rhat |_{\fh
}$. By the Cauchy formula $| (\vhat - \nu '\tau - v')_{\gamma _{\rhat
}} |_{\fh } \le c|\rhat _{\gamma }|_{\fh } $.  Since the $\gamma $
derivative of $\vhat - \nu '\tau - v' $ is bounded by the sum of these
two estimates and since $\nu '$ is independent of $\gamma $, part (i)
is proved.

\Proof{ of (ii) and (iii) } Apply the same argument to $r$ as a
function of $\gamma _{\vhat }$ and $\gamma _{\rhat }$. Part (iii) uses
uniform bounds by $c$ on the disks $\|\gamma _{\rhat } \rhat _{\gamma }
\|_{w^{2},h} \le c_{1}$ and $|\gamma _{\vhat }\vhat _{\gamma }|_{h}
\le c_{1}$.  \qed

\Proof{ of Proposition~\ref{prop:error-for-b}} As in the proof of
Theorem~\ref{prop:beta-lambda-recursion} all parameters are rotated by
$\theta $ in the complex plane, but here we simplify notation by
writing $b, \beta , \lambda , r, \Gamma $ in place of $b (\theta ),
\beta (\theta ) , \lambda (\theta ) , r (\theta ), \Gamma (\theta )$.

\newcommand{\rmainsubgv}{r_{\text{main },\gamma _{v}}}

\emph{Claim:} Define $\fh _{j}$ as before (above
(\ref{eq:r-theta-claim})). There exists $c_{L}$ such that
\begin{align}\label{eq:r-gamma-bound}
&\|r_{j,\gamma }\|_{w,h_{j}} \le c_{L} \|\Gamma (\beta _{j-1})
\|^{\frac{1}{2}}
|b_{j-1}|\nonumber \\
&|r_{j,\gamma }|_{\fh _{j} } \le c_{L} |\lambda _{j}|^{2} \|\Gamma
(\beta _{j-1}) \|^{3}|b_{j-1}|.
\end{align}
By (\ref{eq:rhat-formula}), $\rhat $ is a function of $\gamma _{\rmain
}$ and $\gamma _{v}$ and $\gamma _{r}$, and note that $\rhat _{\gamma
_{\rmain }} = \rmainsubgv $.  By
Lemma~\ref{lem:r-main-estimates}, there are uniform bounds on the
$w^{2},h$ and $\fh $ norms of $\rmain $ in the disk $|\gamma _{\rmain
}||b| \le c h^{-4}$. By the Cauchy estimate and
(\ref{eq:rhat-formula}),
\[
	\|\rmainsubgv \|_{w^{2},h} \le c_{L} \|C \|^{2}|b|,
\hspace{3mm}
	|\rmainsubgv |_{\fh}
\le
	c_{L} \|C \|^{3} |\lambda |^{2} |b|.
\]
We obtain the same estimates on the $\gamma _{v}$ derivative of $\rhat
$ because (\ref{eq:rhat-bound1}, \ref{eq:rhat-bound2}) are uniform on
a disk $|\gamma _{v}| |b| \le c h^{-4}$.

Consider the $\gamma _{r}$ derivative of $\rhat $.  The derivative of
the second term in (\ref{eq:rhat-formula}) is
\[
\cS E_{q-1} (\Delta ) 
(e^{-v})^{\cG _{1}\setminus \{x\}}r_{x,\gamma _{r}} ,
\]
where there is no sum over $x$ and $x=0$ because $r_{x,\gamma _{r}}=0$
unless $x=0$ and then $r_{x,\gamma _{r}}=r_{\gamma _{r}}$. The absence
of the sum over $x$ saves a factor of $L^{4}$ when the argument
leading to (\ref{eq:one-r-term}) is repeated so that
\[
\|
\cS (e^{-v})^{\cG _{1}\setminus \{x\}} \cS r_{x,\gamma _{r}} 
\|_{w^{2}, 2h} 
\le cL^{-6} \|r_{\gamma _{r}}\|_{w,h} ,
\]
and the same factor $L^{4}$ is also saved in estimating the $\fh $
norm.  These terms are the dominant contribution to the $\gamma _{r}$
derivative of $\rhat $: recall the $w^{2},2h$ estimate on the third
term in (\ref{eq:rhat-formula}) in the proof of
Proposition~\ref{prop:control-of-r}. It is uniform on the disk
$\|\gamma _{r}r_{\gamma _{r}} \|_{w,h} \le \|r \|_{w,h}$, so that the
$\gamma _{r}$ derivative of this term is bounded by $ c_{L}\|r\|_{w,h}
\|r_{\gamma _{r}}\|_{w,h}$.  This can be absorbed into a small change
in the constant $c$ in $cL^{-6} \|r_{\gamma _{r}}\|_{w,h}$ because $r$
is small by (\ref{eq:r-theta-claim}). The $\gamma _{r}$ derivative of
the fourth term can also be absorbed because it is down in norm by
$h^{-2q}$ with $q=1$. In summary,
\begin{align*}
&\|\rhat _{\gamma }\|_{w^{2},2h} \le 
c_{L}  \|C \|^{2} |b| + cL^{-6}\|r_{\gamma } \|_{w,h} \nonumber\\
&|\rhat _{\gamma }|_{\fh '} \le 
c_{L}  \|C \|^{3} |\lambda |^{2}|b| + cL^{-6} (\fh '/\fh )^{6} |r_{\gamma } |_{\fh }
+ c_{L}\|C \|^{q}h^{-2q}\|r_{\gamma } \|_{w,h} ,
\end{align*}
where the second equation is obtained in the same way using the $\fh $
norm and $q=5$. By this estimate and parts (ii) and (iii) of
Corollary~\ref{cor:extraction},
\begin{align*}
&\|r_{j+1,\gamma } \|_{w,h_{j+1}} \le c_{L} \|\Gamma (\beta _{j})
\|^{\frac{1}{2}}|b_{j}| +
cL^{-6}\|r_{j,\gamma } \|_{w,h_{j}} \nonumber\\
&|r_{j+1,\gamma }|_{\fh _{j+1}} \le c_{L} \|\Gamma (\beta _{j}) \|^{3}
|\lambda _{j}|^{2} |b_{j}| + c L^{-6}  (\fh _{j+1}/\fh _{j})^{6}
|r_{j,\gamma }|_{\fh _{j}} 
+
c_{L} \|\Gamma _{j}\|^{q} h_{j}^{-2q} \|r_{j,\gamma } \|_{w,h_{j}} ,
\end{align*}
where, in the first estimate, the term $\|\rhat \|_{w^{2},h}|\vhat
_{\gamma }|_{h} $ of Corollary~\ref{cor:extraction} was absorbed by a
change of constant into $c_{L}\|\Gamma (\beta ) \|^{1/2}|b|$, using
Claim $1'$ and $|\lambda ||\vhat _{\gamma }|_{h} \le c|b|$. Likewise, in
the second estimate, the term $|\rhat |_{\fh } |\vhat _{\gamma }|_{\fh
}$ was absorbed into $c_{L} \|C \|^{3} |\lambda |^{2} |b|$.  The claim
(\ref{eq:r-gamma-bound}) follows by induction ({\em c.f.} the proof of
(\ref{eq:r-theta-claim})).

By Corollary~\ref{cor:extraction} (i), Claims 1 and $1'$, and the
estimates above on $\rhat _{\gamma }$,
\[
|{\vhat}_{\gamma } - v'_{\gamma }|_{\fh _{j}} \le c_{L} \|\Gamma _{j-1}
\|^{3}|\lambda _{j-1}|^{2}|b_{j-1}|.
\]
We change the indices $j-1$ to $j$ { on the right} by increasing $c_{L}$.  By the
definition of $\epsilon _{\ast ,j}$ in
Section~\ref{section:greens-function}, $|{\vhat} _{\gamma } -
{ v}'_{\gamma }|_{\fh_{ j} }$ dominates $|\epsilon_{0,j}|$,
$|\epsilon_{1,j}|\fh _{j}^{2}$ and $|\epsilon_{2,j}|\fh
_{j}^{4}$. Solving the inequality for the $\epsilon _{\ast }$
concludes the proof of Proposition~\ref{prop:error-for-b}. \qed


\appendix

\section{Convolution of Forms and Supersymmetry}\label{appendix:form-convolution}
\setcounter{theorem}{0}
\setcounter{equation}{0}

Convolution $f,g \rightarrow \int f(u-v)g(v)\,dv$ of functions has a
natural extension to forms. Furthermore many associated facts such as
the closure of Gaussian functions under convolution also have form
analogues. This is because functions and forms both pull \emph{back}
under the map
\begin{equation}\label{eq:pullback1}
u,v \rightarrow u-v, \hspace{3mm} \CC ^{N}\times \CC ^{N}\rightarrow
\CC ^{N}.
\end{equation}

To make this extension we first must define the integral of a form
over a linear manifold of dimension less than the form. Thus, suppose
$V$ is a complex linear subspace of $\CC ^{m}$ and $\omega $ is
an integrable form on $\CC ^{m}$.  Given any complementary
subspace $V^{\perp}$ we define a form $\int _{V}\omega $ on
$V^{\perp}$ by requiring that
\[
\int _{V^{\perp}} \, \omega ^{\perp} \, \int _{V} \, \omega = \int
_{\CC ^{m}} \omega ^{\perp} \omega 
\]
holds for all $\omega ^{\perp}$ on $V^{\perp}$, where, on the
right-hand side, $\omega ^{\perp}$ is the form on $\CC ^{m}$ obtained
by pulling back the projection from $\CC ^{m}\approx V^{\perp} \oplus
V$ to $ V^{\perp}$.  Taking $V=\CC ^{M}$ as a subspace of $\CC ^{N+M}$
we have the following form analogue of a familiar Gaussian identity:

\begin{proposition}\label{prop:gaussian3} Let $A$ be a matrix on $\CC ^{N+M}
= \CC ^{N} \times \CC ^{M}$ with positive real part. Define the
quadratic form $S_{A}$ as in (\ref{eq:S-sub-A}). Then
\[
\int _{\CC ^{M}} e^{-S_{A}} = e^{-S_{A_{u}}} ,
\]
where $A_{u}$ is the $N \times N$ matrix obtained by inverting 
$(A^{-1})_{ij}$ with indices restricted to $1\le i,j \le N$.
\end{proposition}

The partial integration over \emph{complex linear} subspaces has the
following desirable property: 
\begin{lemma}\label{lem:Q-commutes-with-int} If $\omega $ is a smooth,
rapidly decaying form, then $Q \int _{V} \omega = \int _{V} Q\omega
$.
\end{lemma}
\proof{} 
\[
\int _{V^{\perp}} \omega ^{\perp} \int _{V} Q\omega = \int _{\CC ^{m}}
\omega ^{\perp} Q\omega
\]
Since $Q$ is a derivation and $\int Q (\omega ^{\perp} \omega) =0$,
the right-hand side is $-\int _{\CC ^{m}} ( Q\omega ^{\perp})
\omega $. Since $V$ and $V^{\perp}$ are complex linear, this is the
same as
\[
-\int_{V^{\perp}} ( Q\omega ^{\perp}) \int _{V} \omega =
\int_{V^{\perp}} \omega ^{\perp} Q\int _{V} \omega.
\]
Therefore,
\[
\int _{V^{\perp}} \omega ^{\perp} \int _{V} Q\omega = \int_{V^{\perp}}
\omega ^{\perp} Q\int _{V} \omega.
\]
This proves the claim because $\omega ^{\perp} $ is arbitrary. \qed

\Proof{ of Proposition~\ref{prop:gaussian3}} Let $(u,v) \in \CC ^{N}
\times \CC ^{M}$. $\exp (-S_{A})$ is a Gaussian times a sum of
constant forms. Since Gaussian functions remain Gaussian when
variables are integrated out,
\[
\int _{\CC ^{M}} e^{-S_{A}} = e^{-uA_{u}\bar{u}} \times \text{ some
constant form } \omega.
\]
The covariance of a normalized Gaussian is always the inverse of the
matrix in the exponent. This identifies $A_{u}$ because the covariance
of the $u$ variables is not changed by integrating out $v$.  By
Lemma~\ref{lem:Q-commutes-with-int}, $Q (\exp (-uA_{u}\bar{u} )\omega
)= 0$. Multiply both sides of this equation by $\exp (uA_{u}\bar{u} +
du A_{u} d\bar{u}/ (2\pi i) )$ which is supersymmetric and therefore
commutes past $Q$. Therefore,
\[
Q (e^{du A_{u} d\bar{u}/ (2\pi i)}\omega ) = 0.
\]
Since the only constant forms annihilated by $Q$ are constants, 
\[
\omega = \text{ const } e^{- du A_{u} d\bar{u}/ (2\pi i)}.
\] 
Therefore
\[
\int _{\CC ^{M}} e^{-S_{A}} = \text{ const } e^{-uA_{u}\bar{u} - du A_{u}
d\bar{u}/ (2\pi i)}.
\]
The constant is one because the integral of both sides over $\CC ^{N}$
is one by Lemma~\ref{lem:gaussian1}. \qed 

Next comes the form analogue of the well known closure property of
Gaussian convolutions.  $S_{A} (u-v)$ denotes the pullback of $S_{A}$
by the map (\ref{eq:pullback1}). 
\begin{corollary}\label{cor:gaussian-form-convolution}
\begin{equation*}
\int e^{-S_{A} (u-v)}  e^{-S_{B} (v)} = e^{-S_{C} (u)},
\end{equation*}
with $C^{-1} = A^{-1}+B^{-1}$. The integral is over $v \in \CC ^{N}$. 
\end{corollary}

\proof{} Apply Proposition~\ref{prop:gaussian3} noting that the
left-hand side is a Gaussian form $\exp (-S_{\tilde{A}} )$ with
\[
\tilde{A} =
\left[
\begin{array}{cc}
A, &-A\\
-A, & A+B
\end{array}
\right]
\text{ whose inverse is }
\left[
\begin{array}{cc}
A^{-1}+B^{-1},& B^{-1}\\
B^{-1},&B^{-1}
\end{array}
\right].
\]
\qed 

We conclude with a characterization of supersymmetric forms on $\CC $.

\begin{lemma}\label{lem:supersymmetric-functions} (i) If
$\omega $ is a smooth even supersymmetric form on $\CC $ then $\omega
=f (\tau )$ for some smooth function $f$.  (ii) If $\omega
$ is a smooth odd degree supersymmetric form on $\CC $ then $\omega =
f (\tau ) (\phi \,d\phibar \,+ \phibar \,d\phi )$ for some smooth
function $f$.
\end{lemma}

\proof{} (i) Any even form can be written as $\omega = a +
b \,d\phi \, d\phibar $, where $a,b$ are functions of $\phi $. Then
\begin{align*}
Q\omega &= a_{\phi }\,d\phi + a_{\phibar }\,d\phibar - 
(2\pi i) b \phi \,d\phibar - (2\pi i) b \,d\phi \,\phibar.
\end{align*} 
Therefore supersymmetry implies that the partial derivatives satisfy
$a_{\phi }= 2\pi i \phibar b$, $a_{\phibar }=2\pi i \phi b$ which
implies there is $f$ such that $a = f (\phi \phibar )$ and $b = (2\pi
i)^{-1} f'(\phi \phibar )$.  Therefore $\omega =f (\tau )$. (ii)
Away from $\phi =0$ we can write $\omega = a\phi
\,d\phibar \,+ b \phibar \,d\phi $.  $Q\omega =0$ implies $a=b$ and
then $Qa=0$, so $a=a (\tau )$. \qed

\section{Dirichlet Boundary Conditions}\label{appendix:dirichlet-bc}
\setcounter{theorem}{0}
\setcounter{equation}{0}

\renewcommand{\EE}[1]{\mathbb{E}^{\Lambda }\left(#1\right)} 

$U^{\Lambda ,\beta } (x) = G_{\lambda =0}^{\Lambda} (\beta ,x)$ is the
$\beta $ potential for the hierarchical Levy process $\omega(t)$ killed on
first exit.  Recall that the Hierarchical lattice is invariant under
the map $L^{-1}$ which is the shift $x \longmapsto x +\cG_1$, or
equivalently, when $x = (\dots ,x_{2},x_{1},x_{0})$, $ L^{-1} x =
(\dots ,x_{3},x_{3},x_{1})$.  This map induces a scaling $\cS $ that
acts on Green's functions by $\cS U (x) = L^{-2}U (L^{-1}x)$ and on
$\beta $ by $L^{2}\beta = L^{2}\beta $. We also define $\Lambda /L=
\{\cS x|x \in \Lambda \}$.

The main result for this section is the \emph{scale decomposition}
\begin{equation*}
U^{\Lambda } (\beta ,x) = \cS U^{\Lambda /L} (L^{2}\beta ,x) +
\Gamma(\beta ,x) ,
\end{equation*}
where
\[
\Gamma (\beta,x) = \frac{1}{\gamma +\beta }
(\one _{\cG _{0}} - \frac{1}{n} \one_{\cG _{1}}),
\]
and $\gamma =1$ for the four dimensional hierarchical process. This is
the only property of the hierarchical Levy process that is used in the
renormalization group analysis.  The remainder of this section is
provided for a completeness but plays no further role in the paper.

Recall from \cite{BEI92}, that conditional on jumping, the probability
of jumping from $x$ to $y$ is proportional to $|x|^{-6}$. This jump
law was chosen because it makes the process scale in the same way as
random walk on a four dimensional simple cubic lattice, namely for
$N=\infty $, we have $U^{\beta =0} (x) \propto |x|^{-2}$.  The
\emph{jump rate} $r$ is chosen so that the constant of proportionality
is one.  Equation (\ref{eq:dirichlet-scale-decomp}) is an immediate
consequence of
\begin{proposition}\label{prop:Green1}
\[
U^{\Lambda,\beta } (x) = \sum _{j=0}^{N-1}
\cS ^{j} \Gamma (\cS ^{j}\beta,x) +
(r + \cS ^{N}\beta)^{-1} \cS ^{N} \one _{\cG _{0}} (x)
\]
\end{proposition}

\proof{} We will prove a more general result by considering a unit ball
$\cG _{1}$ with $n$ elements and
\[
q (x-y) = c |x-y|^{-\alpha }, \hspace{2mm} = 0 \text{ if } y = x.
\]
Note that
\[
\int _{\cG_{k}}
q (x) \,dx = 1 - n^{k} L^{-\alpha k}.
\]

By repeating the proof of Lemma~2.2 on page 89 of \cite{BEI92}, taking
into account the killing, we find
\[
\EE{\pair{\omega(t),\xi}} = \exp (-t\psi (\xi )),
\]
with
\[
\psi (\xi ) = 
r - r\int _{\cG _{N}} q (x) \pair{x,\xi}\,dx.
\]
Since $\int q \,dx =1$
\[
\psi (\xi )
= 
r\int _{\cG _{N}} q (x) [1-\pair{x,\xi}]\,dx +
r\int _{\cG \setminus \cG _{N}} q (x) \,dx.
\]

\noindent \emph{Case: } $\xi \in \cH_{N}$.  Recall from \cite{BEI92}
that the dual ball $\cH _{j}$ is defined by: $\xi \in \cH _{j}$ if
$\pair{x,\xi }=1$ for every $x\in \cG _{j}$. Therefore, for $\xi \in
\cH_{N}$, 
\[
\psi (\xi ) = 
r\int_{\cG \setminus \cG _{N}} q (x)\,dx = 
r n^{N} L^{-\alpha N}.
\]

\noindent \emph{Cases: } $\xi \in \cH _{j}\setminus \cH _{j+1}$, with
$j=0,1,\dots N-1$. By the calculation starting at the bottom of page
89 of \cite{BEI92},
\begin{align*}
\int _{\cG _{N}} q (x) [1-\pair{x,\xi}]\,dx 
&=
cn^{j+1}L^{-\alpha (j+1)} + 
c\sum _{k=j+2}^{N}L^{-\alpha k} (n^{k}-n^{k-1}) \\
&=
cn^{j+1}L^{-\alpha (j+1)} + 
\int _{\cG_{N}\setminus \cG _{j+1}} q (x) \,dx.
\end{align*}
Therefore
\[
\psi (\xi ) = 
rcn^{j+1}L^{-\alpha (j+1)} + 
r \int _{\cG \setminus \cG _{j+1}} q (x) \,dx
= \gamma n^{j}L^{-\alpha j}.
\]
where $\gamma = r (1+c) n L^{-\alpha }$ is the same as the $\gamma $
in \cite{BEI92}.

Therefore, using $\one _{\cH _{j}} - \one _{\cH _{j+1}}$ to isolate
these cases:
\begin{align*}
(\beta + \psi (\xi ))^{-1} &= 
\sum _{j=0}^{N-1} (\beta + \gamma n^{j}L^{-\alpha j})^{-1}
(\one _{\cH _{j}} - \one _{\cH _{j+1}}) + 
(\beta + rn^{N}L^{-\alpha N})^{-1} \one _{\cH _{N}}.
\end{align*}
Then we invert the Fourier transform using lemma~2.1 in \cite{BEI92},
\begin{align*}
U^{\Lambda,\beta } &= \sum _{j=0}^{N-1}
(\beta + \gamma n^{j}L^{-\alpha j})^{-1} n^{-j}(\one _{\cG _{j}} - 
\frac{1}{n} \one_{\cG _{j+1}}) +
(\beta + rn^{N}L^{-\alpha N})^{-1} n^{-N} \one _{\cG _{N}}.
\end{align*}

The proposition now follows by choosing $\alpha, n$ so that $n =
L^{\alpha -2}$ and obtain
\[
U^{\Lambda,\beta } (x) = \sum _{j=0}^{N-1}
n^{-j}L^{2j} \ \Gamma (\beta L^{2j},\frac{x}{L^{j}}) +
n^{-N} L^{2N} (r + \beta L^{2N})^{-1}\one _{\cG _{0}}
(\frac{x}{L^{N}}) ,
\]
and then setting $\alpha =6$. \qed 

\remark Choosing $\alpha, n$ so that $n = L^{\alpha -2}$ and $n=L^{d}
$ gives the canonical scaling factor $n^{-j}L^{2j} = L^{- (d-2)}$ so
that the infinite volume potential ($N\rightarrow \infty$) with $\beta
=0$ is
\[
U^{0} (x) =  \sum _{j=0}^{\infty }
L^{- (d-2)j} \ \Gamma (0,\frac{x}{L^{j}}) = \rho |x|^{- (d-2)}
\]
for $x\not =0$ and $d>2$.  We choose $r$ and thereby $\gamma = r
(1+c)n L^{-\alpha }$ so that $\rho =1$. This requires, as in
\cite{BEI92},
\[
\gamma = \frac{1-L^{-2}}{1 - L^{4-\alpha }}, \hspace{4mm}
r = \frac{1-L^{-2}}{1 - L^{4-\alpha }}
\frac{1 - L^{2-\alpha }}{1 - L^{-\alpha }}.
\]


\section{Calculations for Proposition~\ref{prop:recursion1}}
\label{appendix:feynman-diagrams}
\setcounter{theorem}{0}
\setcounter{equation}{0}

\newcommand{\tauvertup}[2]{
\put(#1,#2){
	\put(0,0){\vector(1,1){10}}
	\put(-10,10){\vector(1,-1){10}}
	}
}
\newcommand{\tauvertleft}[2]{
\put(#1,#2){
	\put(0,0){\vector(-1,-1){10}}
	\put(-10,10){\vector(1,-1){10}}
	}
}
\newcommand{\tauvertdown}[2]{
\put(#1,#2){
	\put(0,0){\vector(-1,-1){10}}
	\put(10,-10){\vector(-1,1){10}}
	}
}
\newcommand{\tauvertdownagain}[2]{
\put(#1,#2){
	\put(0,0){\vector(1,-1){10}}
	\put(-10,-10){\vector(1,1){10}}
	}
}
\newcommand{\tauvertright}[2]{
\put(#1,#2){
	\put(0,0){\vector(1,-1){10}}
	\put(10,10){\vector(-1,-1){10}}
	}
}
\newcommand{\tauvertsquared}[2]{
\put(#1,#2){
	\tauvertup{0}{1}
	\tauvertdown{0}{-1}
	}
}
\newcommand{\tauvertsquaredagain}[2]{
\put(#1,#2){
	\tauvertup{0}{1}
	\tauvertdownagain{0}{-1}
	}
}
\newcommand{\tauvertsquaredsideways}[2]{
\put(#1,#2){
	\tauvertleft{-1}{0}
	\tauvertright{1}{0}
	}
}
\newcommand{\leftlaptauvertsquared}[2]{
\put(#1,#2){
	\tauvertsquared{0}{0}
	\qbezier (-10,-11) (-30,0) (-10,11)
	}
}
\newcommand{\rightlaptauvertsquared}[2]{
\put(#1,#2){
	\tauvertsquared{0}{0}
	\qbezier (10,-11) (30,0) (10,11)
	}
}
\newcommand{\toplaptauvertsquared}[2]{
\put(#1,#2){
	\tauvertsquared{0}{0}
	\qbezier (-10,11) (0,11) (10,11)
	}
}
\newcommand{\bottomlaptauvertsquared}[2]{
\put(#1,#2){
	\tauvertsquared{0}{0}
	\qbezier (-10,-11) (0,-11) (10,-11)
	}
}
\newcommand{\rightlaptauvertsquaredsideways}[2]{
\put(#1,#2){
	\tauvertsquaredsideways{0}{0}
	\qbezier (10,-10) (30,0) (10,10)
	}
}

To classify the different algebraic expressions that result from
carrying out the derivatives in the formula in Lemma~\ref{lem:Q} we
introduce the Feynman diagram notation.  The diagrams
\begin{center}
\begin{picture} (200,10)
\tauvertup{10}{0} \tauvertdown{100}{10} \tauvertdownagain{175}{10}
\end{picture}
\end{center}
all represent $\sum  \tau _{x}$.  The incoming vector symbolizes the
$\phi_{x}$ and the $\psi _{x}$ and the outgoing vector symbolizes the
$\phibar _{x}$ and $\bar{\psi }_{x}$. The vectors are called
\emph{legs}. The common vertex signifies the single sum over
$x$ and the sum over a term $\phi_{x}\phibar _{x}$ and a term $\psi
_{x}\bar{\psi }_{x}$, whereas two vertices close together as in the
two diagrams
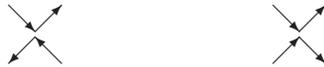
\begin{center}
\begin{picture} (150,20)
\tauvertsquared{10}{10} \tauvertsquaredagain{110}{10}
\end{picture}
\end{center}
represent $\sum \tau ^{2} _{x} $, in which there is a single sum over
$x$ but each $\tau _{x}$ is a sum of two contributions
$\phi_{x}\phibar _{x}$ and $\psi _{x}\bar{\psi }_{x}$. The action of
the Laplacian $\Delta _{\Gamma }$ on $\sum \tau ^{2} _{x}$ is
symbolized by joining an outgoing leg to an incoming leg as in
\begin{equation}\label{eq:vertices1}
\begin{picture}(275,20)
\leftlaptauvertsquared{10}{10} 
\rightlaptauvertsquared{90}{10}
\toplaptauvertsquared{170}{10}
\bottomlaptauvertsquared{250}{10}
\end{picture}
.
\end{equation}
The second two diagrams each contain a closed loop. This closed loop
is a factor of $\Delta _{\Gamma }\tau = 0$ in the algebraic expression
represented by the diagram: the anticommuting $\psi $ derivatives in
$\Delta _{\Gamma }$ give a contribution that exactly cancels the
contribution from the $\phi$ derivatives. Closed loops are always a
factor which is the sum of two canceling contributions so diagrams
containing closed loops will not be exhibited.

We classify all the terms that arise by applying $\Delta _{\Gamma
}^{k}$ by drawing all possible diagrams with $k$ pairs of consistently
oriented legs joined. A joined pair of legs is called a
\emph{line}. Each line is associated to a factor $\Gamma (x,y)$ coming
from $\Delta _{\Gamma }$. Consistently, a line joining legs at the
same vertex carries the factor $\Gamma (x-x)$ from $\Delta _{\Gamma
}$. 

Set $\gamma =0$ so that there is no observable and consider
\[
V = \sum _{y \in x +\cG _{1}}  v_{y}.
\]
By (\ref{eq:vertices1}), $\Delta _{\Gamma }V$ involves two
non-vanishing diagrams, but since both diagrams represent the same
algebraic expression, $\lambda \Gamma (0) \sum \tau _{x} $, we write
one diagram with the combinatoric coefficient $2$,
\begin{center}
\begin{picture}(170,25)

\put (0,10) {$V_{1} := e^{\Delta _{\Gamma }}V = $}

\tauvertsquared{100}{10}

\put (120,10) {$+ \ 2$}

\rightlaptauvertsquared{160}{10} 

\end{picture} 
\hspace{5mm}
.
\end{center}

The graphical representation for $\frac{1}{2} V_{1}\join _{\Gamma }
V_{1}$ is
\begin{center}
\begin{picture}(300,25)

\put (75,10){

	\put (-35,0){$4$}
	\tauvertsquared{-15}{0}
	\qbezier (-5,11) (0,11) (5,11)
	\tauvertsquared{15}{0}
}

\put (160,10){

	\put (-50,0){$+ \ 8$}
	\tauvertsquared{-15}{0}
	\qbezier (-5,11) (0,11) (5,11)
	\rightlaptauvertsquared{15}{0}
}

\put (265,10){

	\put (-60,0){$+ \ 4$}
	\leftlaptauvertsquared{-15}{0}
	\qbezier (-5,11) (0,11) (5,11)
	\rightlaptauvertsquared{15}{0}

}
\end{picture}
.
\end{center}
First diagram: any of 4 legs in left-hand vertex pairs with either of
2 legs in right-hand vertex and prefactor $\frac{1}{2}$.  Second
diagram: any of four legs pairs with one leg, prefactors
$\frac{1}{2} 2$ and a factor $2$ because we can interchange the 4 leg
vertex with the 2 leg vertex.  Third diagram: either of 2 legs can
pair with one leg, prefactors  $\frac{1}{2} 4$.

The graphical representation for $\frac{1}{2} \frac{1}{2!} V_{1}\join
_{\Gamma }^{2}V_{1}$ is
\begin{center}
\begin{picture}(300,25)

\put (35,10){

	\put (-35,0){$2$}
	\tauvertsquaredagain{-15}{0}
	\qbezier (-5,11) (0,11) (5,11)
	\qbezier (-5,-11) (0,-11) (5,-11)
	\tauvertsquaredagain{15}{0}
}

\put (120,10){

	\put (-50,0){$+ \ 2$}
	\tauvertsquared{-15}{0}
	\qbezier (-5,11) (0,11) (5,11)
	\qbezier (-5,-11) (0,-11) (5,-11)
	\tauvertsquared{15}{0}
}

\put (205,10){

	\put (-50,0){$+ \ 4$}
	\tauvertsquared{-15}{0}
	\qbezier (-5,11) (0,11) (4,10)
	\qbezier (-5,-11) (0,-11) (4,-10)
	\tauvertsquaredsideways{15}{0}
}

\put (280,10){

	\put (-50,0){$+ \ 4$}
	\tauvertsquared{-15}{0}
	\qbezier (-5,11) (0,11) (5,11)
	\qbezier (-5,-11) (0,-11) (5,-11)
	\tauvertsquared{15}{0}
	\rightlaptauvertsquared{15}{0}
}
\end{picture}
\hspace{8mm}
.
\end{center}
First diagram: there is already a factor of 4 in the first diagram in
$\frac{1}{2} V_{1}\join _{\Gamma } V_{1}$, there is one way to add
additional line to obtain this topology, prefactor $\frac{1}{2!}$.
Second diagram: same.  Third diagram: there is a factor of 4 in the
second diagram in $\frac{1}{2} V_{1}\join _{\Gamma } V_{1}$, there are
two ways to add additional line to obtain this topology, prefactor
$\frac{1}{2!}$. Fourth diagram: there is a factor of 8 in the third
diagram in $\frac{1}{2} V_{1}\join _{\Gamma } V_{1}$, there is one way
to add additional line to obtain this topology, prefactor
$\frac{1}{2!}$.

The graphical representation for
$\frac{1}{2}\frac{1}{3!}V_{1}\join _{\Gamma }^{3} V_{1}$ is
\begin{center}
\begin{picture}(100,20)
\put (50,10){

	\put (-50,0){$4$}

	\put(-15,1){
	\put(-10,10){\vector(1,-1){10}}
	}

	\put(15,-1){
	\put(0,0){\vector(1,-1){10}}
	}

	\qbezier (-15,1) (0,15) (15,1)
	\qbezier (15,1) (0,0) (-15,-1)
	\qbezier (-15,-1) (0,-15) (15,-1)

}
\end{picture}
.
\end{center}
There is 2 in first diagram in $\frac{1}{2} V_{1}\join _{\Gamma
}^{2}/2 V_{1}$, there are 2 ways to add one more line.  There is no
way to add an additional line to the second diagram in $\frac{1}{2}
V_{1}\join _{\Gamma }^{2}/2 V_{1}$.  There is 4 in third diagram in
$\frac{1}{2} V_{1}\join _{\Gamma }^{2}/2 V_{1}$, there are 2 ways to
add one more line. The total is $12$, there is prefactor $\frac{1}{3}$
in $\frac{1}{3!}$.

For $p>3$,  $V_{1}\join _{\Gamma }^{p} V_{1} = 0$, therefore
\begin{align*}
&
	\cS \frac{1}{2} V_{1}\join _{\Gamma} V_{1}
=
	L^{-2}B_{1}\lambda ^{2} \, \tau _{x}^{3}
=
	0 \text{ because } B_{1} = 0
\\&
	\cS \frac{1}{2} \frac{1}{2!} V_{1}\join _{\Gamma }^{2} V_{1}
=
	8B_{2}\lambda ^{2} \, \tau _{x}^{2} +
	4B_{2}B_{0}L^{2}\lambda ^{2} \tau _{x}
\\&
	\cS \frac{1}{2} \frac{1}{3!} V_{1}\join _{\Gamma }^{3} V_{1}
=
	4 B_{3}L^{2} \lambda ^{2} \tau _{x} ,
\end{align*}
so that
\begin{align*}
	\lambdat
&=
	\lambda 
	- 8 B_{2} \lambda ^{2}\\
	\betat
&=
	L^{2}\beta +
	2 B_{0}L^{2}\lambda -
	4L^{2}B_{2}B_{0}\lambda ^{2} -
	4L^{2}B_{3}\lambda ^{2}.
\end{align*}
There are three cases to consider for the observable.  Case (1) there
have been fewer than $N (x)-1$ iterations so that the observable is
$-\gamma b_{1,j}\phi _{0}\phibar _{x_{j}}$ with $|x_{j}| >L$.  Here we
know by Lemma~\ref{lem:no-field-strength} that $b_{i,j}=L^{-2j}$ and
$b_{1,j}=b_{2,j}=0$ and there is no need to calculate anything. Case
(2) $j\ge N (x)$.  For this case $v_{0}$ contains the additional terms
\[
- \gamma (b_{0}+b_{1}\phi _{0}\phibar _{0} + b_{2}\tau _{0}\phi
_{0}\phibar _{0} + b_{3}\tau _{0}).
\]
Therefore in $\cS V_{1}$ we have the additional terms
\[
-b_{0} - b_{1} L^{-2} \phi \phibar - 
b_{2}L^{-4} \tau \phi \phibar - \Gamma
(0)b_{1} - O (\Gamma ^{2})b_{2} - 2\Gamma (0)b_{2}\phi \phibar ,
\]
where we have omitted $0$ and $j$ subscripts.  Terms of the form
$b\tau $ have been omitted for reasons explained at the end of
Section~\ref{section:coordinates}. In the second order part $\cS
Q_{1}$ we have additional terms
\[
O (\Gamma ^{3})\lambda b_{1} + O (\Gamma ^{4})\lambda b_{2} + 
\left(
O (\Gamma ^{2})\lambda b_{1} + O (\Gamma ^{3})\lambda b_{2}
\right)
L^{-2}\phi \phibar + 
O (\Gamma ^{2})\lambda b_{2} L^{-4}\phi \phibar \tau.
\]
There is no $O(\Gamma )\lambda b_{1}$ contribution to $b_{2}$ because
$B_{1}=0$.

Case (3) $j=N (x)-1$. This is almost the same as case (2), but with
$b_{2,j}$ vanishing.


\end{document}